\documentclass[iop]{emulateapj}

\usepackage{amsmath}

\shorttitle{$\Lambda$CDM Cosmology for Astronomers}
\shortauthors{Condon \& Matthews}

\begin{document}

\title{${\Lambda}$CDM Cosmology for Astronomers} 

\author{J.~J.~Condon\altaffilmark{1}}
\affiliation{National Radio Astronomy Observatory,
520 Edgemont Road, Charlottesville, VA 22903, USA}
\email{jcondon@nrao.edu}
\and
\author{A.~M.~Matthews}
\affiliation{Department of Astronomy, University of Virginia,
Charlottesville, VA 22904, USA}
\email{amm4ws@virginia.edu}

\altaffiltext{1}{The National Radio Astronomy Observatory is a facility
of the National Science Foundation operated under cooperative agreement
by Associated Universities, Inc.}

\begin{abstract}
  \noindent
  The homogeneous, isotropic, and flat $\Lambda$CDM universe favored
  by observations of the cosmic microwave background can be described
  using only Euclidean geometry, locally correct Newtonian mechanics,
  and the basic postulates of special and general relativity.  We
  present simple derivations of the most useful equations connecting
  astronomical observables (redshift, flux density, angular diameter,
  brightness, local space density, \dots) with the corresponding
  intrinsic properties of distant sources (lookback time, distance,
  spectral luminosity, linear size, specific intensity, source counts,
  \dots). We also present an analytic equation for lookback time that
  is accurate within 0.1\% for all redshifts $z$.  The exact equation
  for comoving distance is an elliptic integral that must be evaluated
  numerically, but we found a simple approximation with errors $<
  0.2$\% for all redshifts up to $z \approx 50$.\newline
\end{abstract}


\section{Introduction}

{\it ``Innocent, light-minded men, who think that
astronomy can be learnt by looking at the stars without knowledge
of mathematics will, in the next life, be birds.''}---\rm{Plato, Timaeus}
\vskip 4pt

\begin{figure}[!ht]
\includegraphics[trim = 5 0 0 0,scale=0.7]{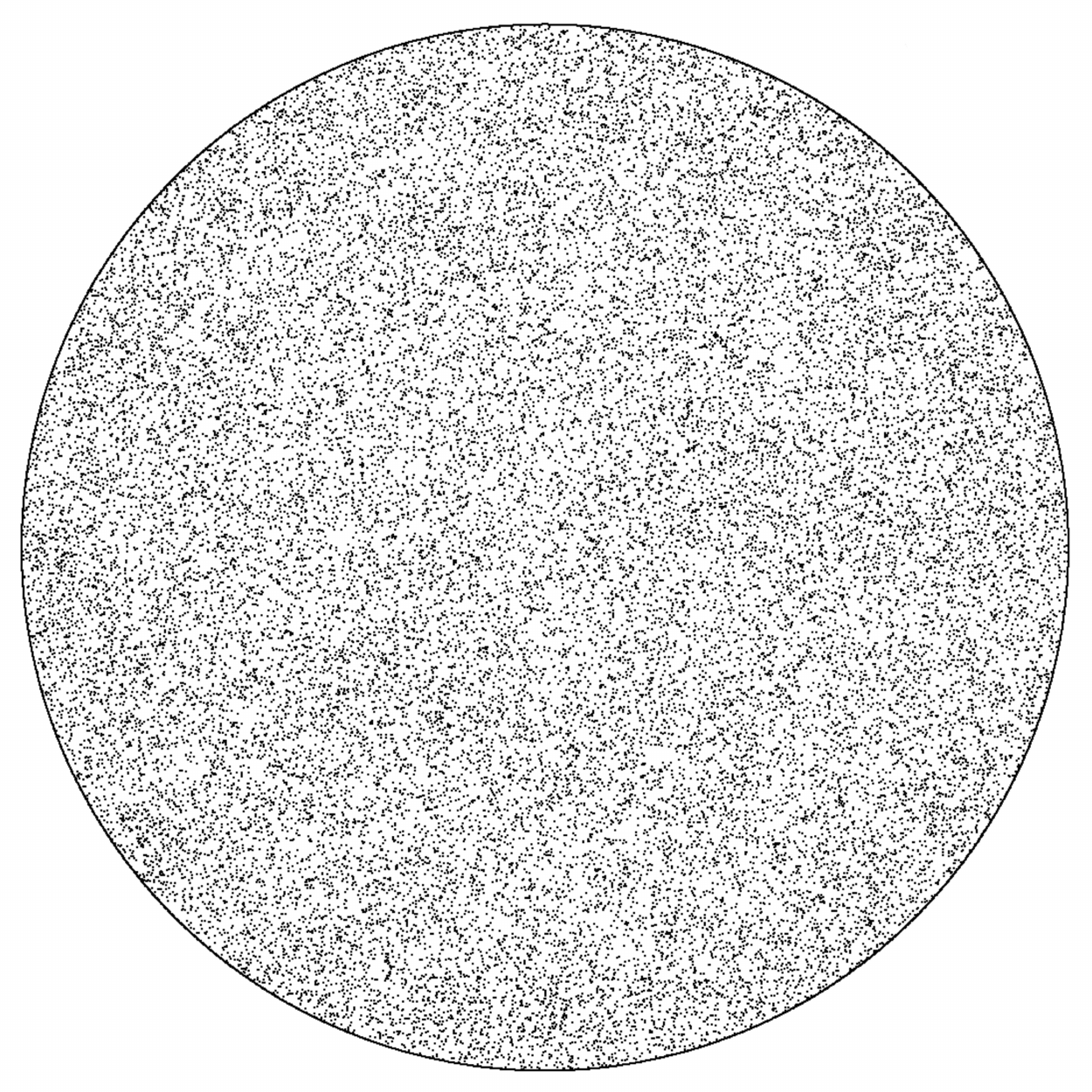} 
\caption{Positions of the $N \sim 4 \times 10^4$ radio sources
  stronger than $S =2.5 \mathrm{~mJy}$ at 1.4~GHz are indicated by
  points on this equal-area plot covering the sky within $15^\circ$ of
  the north celestial pole.  Nearly all of these sources are
  extragalactic and so distant (median redshift $\langle z \rangle
  \sim 1$) that their distribution is quite isotropic.
\label{fig:isotropyncp}}
\end{figure}

According to the cosmological principle and confirmed by observation
(e.g., Figure~\ref{fig:isotropyncp}), the universe is isotropic and
homogeneous on large scales.  It began with a very dense ``big bang''
and has been expanding uniformly ever since.  General relativity can
describe both the geometry and expansion dynamics of the universe.
However, general relativity permits spatially curved universes which
are mathematically complicated.  Extragalactic astronomers were once
faced with the choice of learning general relativity or applying
published relativistic results without really understanding them, at
the risk of being birds in the next life.

In today's ${\Lambda}$CDM ($\Lambda$ for dark energy with constant
energy density and CDM for cold dark matter) concordance model, the
universe is spatially ``flat,'' so its geometry is Euclidian, its
expansion is not affected by curvature, and locally correct
Newtonian calculations can be extended to cosmological scales.
Fortunately for non-mathematicians, flatness allows simple (no
tensors) derivations of accurate equations for the kinematics and
dynamics of cosmic expansion that an undergraduate physics major can
understand.  Such derivations are presented in
Sections~\ref{sec:kinematics} and \ref{sec:dynamics}, and the main
results used by observational astronomers are developed in
Section~\ref{sec:results}.

See David Hogg's useful ``cheat sheet'' \citep{hog99} listing
results from relativistic models with nonzero curvature, and the books
by \citet{pee93} and \citet{wei72} for their derivations.  The
astropy.cosmology Python package at {\tt
  http://docs.astropy.org/\allowbreak en/\allowbreak
  stable/\allowbreak cosmology/} contains utilities for calculating
many of the quantities discussed in this paper.

We use the subscript 0 to distinguish present (redshift zero) values
of evolving quantities.  Unless otherwise noted, all numerical results
are based on a Hubble constant $H_0 = 70 \mathrm{~km~s}^{-1}
\mathrm{~Mpc}^{-1}$ so $h \equiv H_0 /
(100\mathrm{~km~s}^{-1}\mathrm{~Mpc}^{-1}) = 0.7$, plus the following
normalized densities at redshift zero: total $\Omega_0 = 1$, baryonic
and cold dark matter $\Omega_{0,\mathrm{m}} = 0.3$, radiation and
relic neutrinos $\Omega_{0,\mathrm{r}} = h^{-2} \cdot 4.2 \times
10^{-5}$, and dark energy $\Omega_{0,\Lambda} = 1 -
(\Omega_{0,\mathrm{m}} + \Omega_{0,\mathrm{r}}) \approx 0.7$. Note
that many authors just write $\Omega$, $\Omega_\mathrm{m}$,
$\Omega_\mathrm{r}$, and $\Omega_\Lambda$ without the subscript 0 to
indicate the present values of these densities.  Also, $1
\mathrm{~Mpc} \approx 3.0857 \times 10^{19} \mathrm{~km}$ and $1
\mathrm{~yr} \approx 3.1557 \times 10^7 \mathrm{~s}$.

\section{Expansion Kinematics}\label{sec:kinematics}

According to the equivalence principle, all fundamental observers
locally at rest relative to their surroundings anywhere in the
isotropically expanding (and hence homogeneous) universe are in
inertial frames, and their clocks all agree on the time $t$ elapsed
since the big bang. This universal time $t$ is sometimes called world
time or cosmic time, and it equals the proper time of all fundamental
observers.  Fundamental observers didn't have to be present at the
creation to synchronize their clocks; the temperature of the cosmic
microwave background (CMB) radiation is a suitable proxy for time.
Likewise, the CMB appears isotropic only to fundamental observers, and
others can use the CMB dipole anisotropy \citep{kog93} to deduce their
usually small ($v^2 \ll c^2$, where $c \approx 299792
\mathrm{~km~s}^{-1}$ is the vacuum speed of light) peculiar velocities
and correct for them if necessary.

\begin{figure}[!ht]
\includegraphics[trim = 130 360 0 350,scale=0.7]{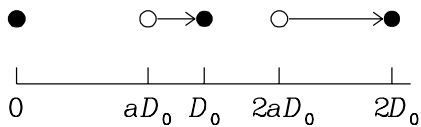} 
\caption{To preserve homogeneity and isotropy, all distances $D = a D_0$,
  $2D = 2 a D_0$,\dots \,between fundamental observers must grow in
  proportion to the universal scale factor $a(t)$.
\label{fig:zoom}}
\end{figure}

Homogeneity and isotropy are preserved if and
only if the small proper distance $D(t)$ between any close pair
of observers expands as
\begin{equation}\label{eqn:enlarge}
  D(t) =  a(t)\,D_0~,
\end{equation}
where $D_0$ is the proper distance now and $a(t)$ is the universal
(meaning, it is the same at every position in the universe)
dimensionless scale factor that grew with time from $a \approx 0$ just
after the big bang to $a(t_0) \equiv 1$ today (Figure~\ref{fig:zoom}).
Equation~\ref{eqn:enlarge} applies to any expansion that preserves
homogeneity and isotropy---the separations of dots on a photo being
enlarged, for example.  The cosmological expansion affects only to the
separations of non-interacting objects, and the dots representing
rigid rulers, gravitationally bound galaxies, etc. do not expand with
the universe.

From Figure~\ref{fig:zoom} it is clear that
\begin{equation}
  \frac {d \ln D}{dt} = \frac{1}{D}\, \frac {dD} {dt} =
  \frac {1} {a D_0} \frac {D_0 \,da} {~~~~dt}  = \frac {\dot{a}} {a} \equiv H(t)
\end{equation}
can vary with time but not with position in space.  $H(t)$
is called the Hubble parameter, and its current value is the Hubble
constant $H_0$.  The time derivative $\dot{D}$ of $D$ in
Equation~\ref{eqn:enlarge} defines the recession velocity of the
nearby observer:
 \begin{equation}\label{eqn:vr}
   v_\mathrm{r} \equiv \dot{D} = \dot{a}\,D_0  =
   \biggl( \frac {\dot{a}}{a} \biggr)D = HD~.
 \end{equation}

Successive wave crests of light emitted with frequency $\nu_\mathrm{e}$
and wavelength $\lambda_\mathrm{e} = c / \nu_\mathrm{e}$ are separated
in time by $dt = \nu_\mathrm{e}^{-1}$ in the source frame.  If the
source is receding from the observer with velocity $v_\mathrm{r} \ll
c$, successive waves must travel an extra distance $v_\mathrm{r} \,dt
= v_\mathrm{r} / \nu_\mathrm{e}$, so their observed wavelength is
\begin{equation}
  \lambda_\mathrm{o}  = \frac {c} {\nu_\mathrm{e}} +
  \frac {v_\mathrm{r}} {\nu_\mathrm{e}}
  = \lambda_\mathrm{e} +  \lambda_\mathrm{e}
  \biggl(\frac {v_\mathrm{r}} {c} \biggr)
  \quad (v_\mathrm{r} \ll c)~
\end{equation}
and $v_\mathrm{r}$ is measurable via the first-order Doppler shift:
\begin{equation}\label{eqn:vrc}
  \frac  {v_\mathrm{r}} {c} =
  \frac {\lambda_\mathrm{o} - \lambda_\mathrm{e}}
  {\lambda_\mathrm{e}} \quad (v_\mathrm{r} \ll c)~.
\end{equation}         
The  redshift $z$ of a source is defined by
\begin{equation}
    z \equiv \frac {\lambda_\mathrm{o} - \lambda_\mathrm{e}}
    {\lambda_\mathrm{e}}~,
\end{equation}
and the domain of this definition extends to all $z$.  Note that most
recession ``velocities'' reported by astronomers are actually
$v_\mathrm{r} = cz$ and may be much larger than the vacuum speed of
light.

Combining Equations~\ref{eqn:vr} and \ref{eqn:vrc} for a
nearby  source at distance $D = c \Delta t$ gives
 \begin{equation}\label{eqn:dotlambda}
   \frac {v_\mathrm{r}} {c} = \frac {\Delta\lambda} {\lambda}
   = \frac {D} {c} \frac {\dot{a}} {a} = \Delta t \, \frac {\dot{a}} {a} =
   \frac{\Delta a} {a} \quad (v_\mathrm{r} \ll c)~.
 \end{equation}
Integrating the local Equation~\ref{eqn:dotlambda} over time:
\begin{equation}
  \int_{\lambda_\mathrm{e}}^{\lambda_\mathrm{o}} \frac{d \lambda}{\lambda} =
      \int_a^1 \frac{da}{a}
\end{equation}
gives the global result that $\lambda_\mathrm{o} / \lambda_\mathrm{e}
= 1 /a $ and $\nu_\mathrm{o} / \nu_\mathrm{e} = a$. Thus the
observable redshift $z$ of any distant source can be used to calculate
the scale factor $a$ of the universe when it emitted the light
seen today:
\begin{equation}\label{eqn:az}
  \boxed{
  a = (1+z)^{-1} }~.
\end{equation}

Because $1 \leq (1+z) < \infty$ is the reciprocal of the scale factor
$0 < a \leq 1$, at high redshifts both $z$ and $(1+z)$ are very
nonlinear and potentially misleading functions of fundamental
quantities such as lookback time (Section~\ref{sec:times}).  Had
astronomers always been able to measure accurate frequency ratios
$\nu_\mathrm{o}/\nu_\mathrm{e} = a$ instead of just small differential
wavelengths $(\lambda_\mathrm{o} -
\lambda_\mathrm{e})/\lambda_\mathrm{e} = z$, most cosmological
equations and results would probably be presented in terms of $a$
today.

The dimension of $H$ is inverse time, but
astronomers originally measured the Hubble constant as the mean
ratio $v_\mathrm{r}/D$ of nearby galaxies, so it is usually written
in mixed units of length and time:
\begin{equation}
  H_0 =  100 \, h \mathrm{\,km\,s}^{-1} \mathrm{~Mpc}^{-1}~.
\end{equation}
Isolating the dimensionless factor $h$ makes it easy to compare
results based on different measured values of $H_0$.  The most recent
measurements of $h$ range from the low $h = 0.669 \pm 0.006$
\citep{pla16} derived by comparing the observed angular power spectrum
of CMB fluctuations with a flat $\Lambda$CDM cosmological model to the
high $h = 0.732 \pm 0.017$ \citep{rie16} based on relatively local
measurements of Cepheid variable stars and Type Ia supernovae used as
standard candles.  The  $3 \sigma$ ``tension'' between these
results is a topic of current research \citep{fre17}.

The Hubble time is defined by
\begin{equation}
  t_\mathrm{H} \equiv H^{-1}~,
\end{equation}
and its present value
\begin{equation}\label{eqn:th0}
  t_\mathrm{H_0} \equiv H_0^{-1} \approx  9.778 \times 10^9 \, h^{-1}
  \mathrm{~yr}  \approx 14.0 \mathrm{~Gyr}
\end{equation}
is a convenient unit of time comparable with the present age of the
universe $t_0$.  Likewise, the current Hubble distance
\begin{equation}\label{eqn:dhub}
  D_\mathrm{H_0} \equiv \frac{c}{H_0} \approx 2998 \, h^{-1}
    \mathrm{~Mpc} \approx 4280 \mathrm{~Mpc}
\end{equation} is a
distance comparable with the present radius of the observable universe.

The lookback time $t_\mathrm{L}(z)$ to a source at any redshift $z$ is
the time photons needed to travel with speed $c$ from the source to the
observer at $z = 0$.  In a homogeneous universe, this global
quantitity is just the sum of the small locally measured proper times $dt$.
In terms of the scale factor $a$ and $H = d \ln (a) / d t$, it is 
\begin{align}
  t_\mathrm{L} &  = \int_t^{t_0} dt' = \int_a^1
     \biggl( \frac {dt'}{d \,\ln(a')} \biggr) d\,\ln(a') \\
  t_\mathrm{L}& = \int_a^1 \frac {da'}{a'H(a')}~.
\end{align}    
The lookback time is usually written in terms of $z$:
\begin{equation}
  t_\mathrm{L} = \int_t^{t_0} dt' =
  \int_z^0 \biggl( \frac {dt'} {dz'}\biggr) \, dz'~.
\end{equation}
To calculate $dt/dz$, take the time derivative of
Equation~\ref{eqn:az}:
\begin{equation}\label{eqn:dzdt}
  \frac {dz} {dt} = \frac{-1} {a^2} \frac {da} {dt} =
  -\frac {1} {a} \biggl( \frac {\dot{a}} {a} \biggr) =
  -(1+z)H
\end{equation}
so
\begin{equation}\label{eqn:tl}
  \boxed{
  t_\mathrm{L} = \int_0^z \frac {dz'} {(1+z') H(z')} }~.
\end{equation}
The dynamical equation specifying the evolution of $H$
is derived in Section~\ref{sec:dynamics}.

\section{Expansion Dynamics}\label{sec:dynamics}

\begin{figure}[!ht]
\includegraphics[trim = 145 300 110 195,scale=0.7]{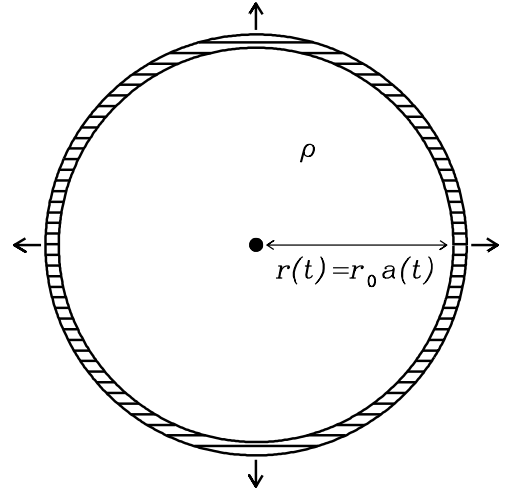} 
\caption{Expansion of the spherical shell with radius $r(t)$ and mean
  density $\rho$ centered on any fundamental observer is affected only
  by the enclosed gravitational mass $M = E/c^2 = (4 \pi r^3 / 3)
  \,\rho$.
\label{fig:shell}}
\end{figure}

The expansion of any small spherical shell with radius $r(t) = r_0
\,a(t) \ll D_\mathrm{H_0}$ and mean density $\rho$ centered on any
fundamental observer is Newtonian in our flat universe because (1) the
net gravitational effect of the surrounding isotropic universe is
zero, in both Newtonian and general relativistic mechanics
\citep{bir23}, and (2) there is no relativistic curvature
acceleration.  The only relativistic results needed are (1) the mean
gravitational mass density $\rho$ is the total relativistic mass
density $E/c^2$, not just the Newtonian rest mass density,
and (2) in a flat universe, the actual mean density $\rho$ always
equals the critical density $\rho_\mathrm{c}$ for which the sum of the
kinetic and gravitational potential energies per unit mass of the
shell is zero:
\begin{equation}
\frac{\dot{r}^2}{2} -  \frac{4 \pi G \rho  r^3} {3r}  = 0~, 
\end{equation}
where $G \approx 6.674 \times 10^{-11} \mathrm{~m}^3 \mathrm{~kg}^{-1}
\mathrm{~s}^{-2}$ is Newton's gravitational constant.  The actual
radius $r_0$ in $r = r_0 \, a(t)$ cancels out, leaving an equation for
the scale factor $a$
\begin{equation}
  \frac{\dot{a}^2}{2} - \frac{4 \pi G\rho a^2}{3} = 0~
\end{equation}
which can be solved for the Hubble parameter
\begin{equation}\label{eqn:Friedmann}
  H^2 = \biggl(  \frac {\dot{a}}{a} \biggr)^2 =
  \frac {8 \pi G \rho}{3}~.
\end{equation} 
Equation~\ref{eqn:Friedmann} is the same as the equation Friedmann
derived from general relativity for a homogeneous, isotropic, and flat
universe.

The present mean density of the flat universe is
\begin{align}\label{eqn:rho0}
  \rho_0 =
  \frac {3 H_0^2}{8 \pi G} 
&  \approx  1.878\times10^{-26}\,h^2 \mathrm{\,kg\,m}^{-3} \nonumber \\
&  \approx 9.20 \times 10^{-27} \mathrm{\,kg\,m}^{-3} ~.\quad~~
\end{align}

The normalized density parameter $\Omega$ is defined as
\begin{equation}\label{eqn:Omegadef}
  \Omega \equiv \frac {\rho}{\rho_\mathrm{c}}~,
\end{equation}
and $\Omega = 1$ for all time in a flat universe. There are three
dynamically distinct contributors to $\Omega$: (1) matter consisting
of ordinary baryonic matter plus cold dark matter particles whose rest
mass nearly equals their total mass, (2) dark energy whose density is
constant, and (3) radiation, primarily CMB photons plus the cosmic
neutrino background (C$\nu$B) of relic neutrinos from the big bang.

\citet{pla16} observations of the CMB angular power spectrum indicate
that $\Omega_0 = \Omega_{0,\mathrm{m}} + \Omega_{0,\Lambda} +
\Omega_{0,\mathrm{r}} = 1.0023 \pm 0.0055$, $\Omega_{0,\mathrm{m}} =
0.315 \pm 0.013$, and $\Omega_{0,\Lambda} = 0.685 \pm 0.013$.
Blackbody radiation at temperature $T$ has energy density $U = 4
\sigma T^4 / c$, where $4 \sigma / c \approx 7.566 \times 10^{-16}
\mathrm{~J~m}^{-3}$ is the radiation constant.  The $T_0 \approx 2.73
\mathrm{\,K}$ CMB has energy density $U_0 \approx 4.20 \times 10^{-14}
\mathrm{~J~m}^{-3}$ and gravitational mass density $\rho_{0,r} = U_0 /
c^2 \approx 4.67 \times 10^{-31} \mathrm{~kg~m}^{-3}$. Massless relic
neutrinos have energy density $U_0 \approx 2.86 \times 10^{-14}
\mathrm{~J~m}^{-3}$ and gravitational mass density $\rho_{0,\nu}
\approx 3.18 \times 10^{-31} \mathrm{~kg~m}^{-3}$ \citep{pee93}. The
total from photons plus neutrinos is $\Omega_{0,\mathrm{r}} \approx
 4.2 \times 10^{-5} \,h^{-2} \approx 8.6 \times 10^{-5}$.

\begin{figure}[!ht]
\includegraphics[trim = 125 280 0 270,scale=0.72]{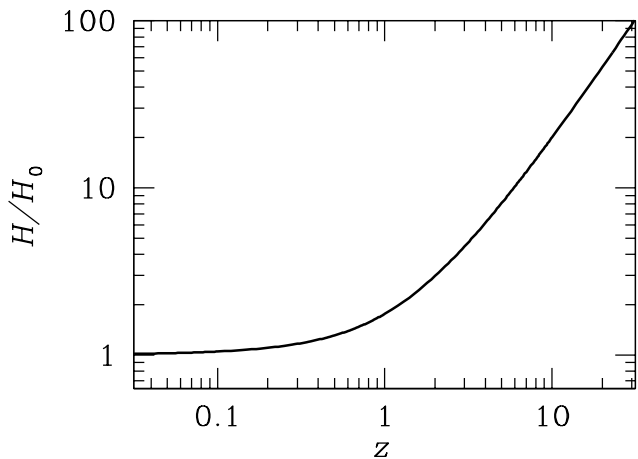} 
\caption{The normalized Hubble parameter $H / H_0$ is nearly constant
  after $(1+z) \lesssim (\Omega_{0,\Lambda} /
  \Omega_{0,\mathrm{m}})^{1/3} \approx 1.33$ and $\rho_\Lambda >
  \rho_\mathrm{m}$.  $H / H_0 \propto (1+z)^{3/2}$ at higher redshifts
  when $\rho_\mathrm{m} > \rho_\Lambda$, and $H/H_0 \propto (1+z)^2$
  at the highest redshifts $z > z_\mathrm{eq} \sim 3500$ when
  $\rho_\mathrm{r} > \rho_\mathrm{m}$.
\label{fig:hofz}}
\end{figure}

Mass conservation of non-relativistic matter implies $ \rho_\mathrm{m}
\propto a^{-3} = (1+z)^3$. In the $\Lambda$CDM model, dark energy is
assumed to behave 
like a cosmological constant:  $\rho_\Lambda \propto a^0 =
(1+z)^0$. The density of radiation (and massless neutrinos) scales
as $\rho_\mathrm{r} \propto a^{-4} = (1+z)^4$ because the number
density of photons is $\propto a^{-3} = (1+z)^3$ and the mass $E/c^2 =
h \nu/c^2$ of each photon scales as $E \propto \lambda^{-1}
\propto (1+z)^1 \propto a^{-1}$.  Inserting these results into Equations
~\ref{eqn:Friedmann}, \ref{eqn:rho0}, and \ref{eqn:Omegadef} leads to
\begin{equation}
  \frac {\rho}{\rho_0} = \frac {H^2}{H_0^2} =
  \frac {\Omega_{0,\mathrm{m}}}{a^3} + \frac {\Omega_{0,\Lambda}}{a^0}
  + \frac {\Omega_{0,\mathrm{r}}}{a^4}~.
\end{equation}
Using Equation~\ref{eqn:az} to replace the scale factor $a$ by the
observable $(1+z)^{-1}$ yields the dynamical equation specifying the
evolution of $H$:
\begin{equation}\label{eqn:Heqn}
  \boxed{ \frac{H}{H_0} = [\Omega_{0,\mathrm{m}} (1+z)^3 +
      \Omega_{0,\Lambda}
      + \Omega_{0,\mathrm{r}} (1+z)^4]^{1/2}}~.
\end{equation}
The symbol
\begin{equation}\label{eqn:edef}
  \boxed{
    E(z) \equiv [\Omega_{0,\mathrm{m}} (1+z)^3 + \Omega_{0,\Lambda}
      + \Omega_{0,\mathrm{r}} (1+z)^4]^{1/2} }
\end{equation}
is a convenient shorthand for subsequent calculations.
Figure~\ref{fig:hofz} shows $H/H_0 = E(z)$ as a function of $z$
for $\Omega_{0,\mathrm{m}} = 0.3$.

The densities of radiation and matter were equal when
$\Omega_{0,\mathrm{r}} (1 + z_\mathrm{eq})^4 = \Omega_{0,\mathrm{m}}
(1+z_\mathrm{eq})^3$ at $z_\mathrm{eq} = (\Omega_{0,\mathrm{m}} /
\Omega_{0,\mathrm{r}}) - 1 \approx 3500$.  The density of matter fell
below that of dark energy at $z \approx (\Omega_{0,\Lambda} /
\Omega_{0,\mathrm{m}})^{1/3} - 1 \approx 0.33$ about 4~Gyr ago.  In
the distant future completely dominated by dark energy, the Hubble
parameter will asymptotically approach $H = H_0
\,\Omega_{0,\Lambda}^{1/2} \approx  84 \,h \mathrm{~km~s}^{-1}
\mathrm{~Mpc}^{-1} \approx 59 \mathrm{~km~s}^{-1} \mathrm{~Mpc}^{-1}$
and the scale factor will grow exponentially [$a \propto \exp(H t)$]
with a time scale $H^{-1} \approx 1.17 \times 10^{10} \, h^{-1}
\mathrm{~yr} \approx 17 \mathrm{~Gyr}$.

\section{Results}\label{sec:results}

\subsection{Cosmic Times}\label{sec:times}

The lookback time $t_\mathrm{L}(z)$ to a source at redshift $z$
can be calculated by inserting Equations~\ref{eqn:Heqn} and
\ref{eqn:edef} into Equation~\ref{eqn:tl}:
\begin{equation}\label{eqn:tlookback}
\boxed{
 \frac {t_\mathrm{L}} {t_\mathrm{H_0}} =
  \int_0^z \frac {dz'}{(1+z')E(z')}}~.
\end{equation}
This integral cannot be expressed in terms of elementary functions,
but the integrand is smooth enough that the integral can be evaluated
numerically by Simpson's rule, at least for finite $z$.

\begin{figure}[!ht]
\includegraphics[trim = 140 280 0 270,scale=0.75]{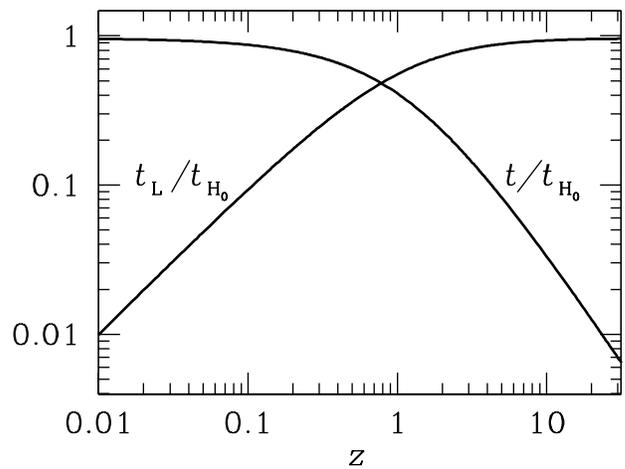} 
\caption{The normalized lookback time $(t_\mathrm{L} / t_{\mathrm{H}_0})$
  and the normalized age $(t/t_{\mathrm{H}_0})$ in a flat $\Lambda$CDM
  universe  with $\Omega_{0,\mathrm{m}} = 0.3$.
\label{fig:tofz}}
\end{figure}

The proper age of the universe $t(z)$ at redshift $z$ is better
evaluated in terms of $a = (1+z)^{-1}$:
\begin{equation}
\boxed{
\frac {t} {t_\mathrm{H_0}} =  \int_0^{(1+z)^{-1}}
\mskip-20mu \frac {a \, da}
{(\Omega_{0,\mathrm{m}} a + \Omega_{0,\Lambda} a^4 +
\Omega_{0,\mathrm{r}})^{1/2}} }~.
\end{equation}
The ratios $t_\mathrm{L} / t_\mathrm{H_0}$ and $ t / t_\mathrm{H_0}$
are plotted as functions of redshift in Figure~\ref{fig:tofz}.  The
present age of the $\Lambda$CDM universe starting with the big bang
($z = \infty$) is $t_0 \approx 0.964\, t_\mathrm{H_0} \approx 
9.42 \times 10^9 \,h^{-1} \mathrm{~yr} \approx 13.5 \mathrm{~Gyr}$ for
$\Omega_{0,\mathrm{m}} = 0.3$.

\begin{figure}[!ht]
\includegraphics[trim = 125 270 0 265,scale=0.72]{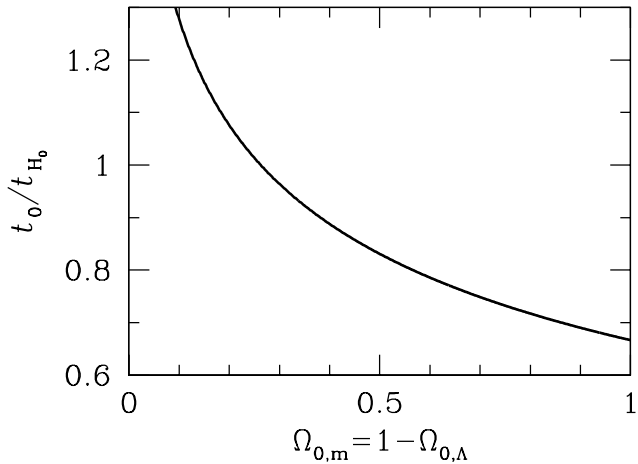} 
\caption{The normalized age  of the universe $t_0 / t_{\mathrm{H}_0}$
as a function of $\Omega_{0,\mathrm{m}} = 1 - \Omega_{0,\Lambda}$
\label{fig:ages}}
\end{figure}

Figure~\ref{fig:tofz} shows that redshifts $z \gg 1$ contribute little
to the age of the universe, so extremely good analytic approximations
to $t_\mathrm{L}/t_{\mathrm{H}_0}$, $t/t_{\mathrm{H}_0}$, and
$t_0/t_{\mathrm{H}_0}$ can be made by ignoring the radiation term
$\Omega_{0,\mathrm{r}}(1+z)^4$ that dominates $E(z)$ only during the
brief period when $z > z_\mathrm{eq} \sim 3500$ ($t / t_{\mathrm{H}_0}
\sim 3.4 \times 10^{-6}$, or $t \sim  5 \times 10^4\,h^{-1}
\mathrm{~yr} \sim 7 \times 10^4 \mathrm{~yr}$).  This simplifies
Equation~\ref{eqn:tlookback} to
\begin{equation}\label{eqn:tapp}
 \frac {t_\mathrm{L}} {t_{\mathrm{H}_0}} \approx \int_0^z 
\frac {dz'} {(1+z') [\Omega_{0,\mathrm{m}} (1+z')^3 +
\Omega_{0,\Lambda}]^{1/2}}~,
\end{equation}
which can be integrated analytically.  See Appendix~\ref{app:tlkbk}
for analytic approximations to $t_\mathrm{L} / t_{\mathrm{H}_0}$
(Equation~\ref{eqn:tlapprox}) and $t / t_{\mathrm{H}_0}$
(Equation~\ref{eqn:tthapprox}).  The current age of the universe normalized by
the Hubble time is very nearly
\begin{equation}\label{eqn:ages}
\boxed{  \frac {t_0} {t_{\mathrm{H}_0}} \approx \frac {2}{3 \,\Omega_{0,\Lambda}^{1/2}}
  \ln \Biggl[ \frac {1 + \Omega_{0,\Lambda}^{1/2}}
    {(1 - \Omega_{0,\Lambda})^{1/2}} \Biggr]}~.
\end{equation}
Figure~\ref{fig:ages} plots $t_0 / t_{\mathrm{H}_0}$ from
Equation~\ref{eqn:ages} as a function of the matter density parameter
$\Omega_{0,\mathrm{m}} = 1 - \Omega_{0,\Lambda}$.  In the limit
$\Omega_{0,\mathrm{m}} = 1 - \Omega_{0,\Lambda} \rightarrow 0$, the
universe would expand exponentially and $t_0 / t_{\mathrm{H}_0}$
would diverge.

\begin{figure}[!ht]
\includegraphics[trim = -30 190 0 240,scale=0.45]{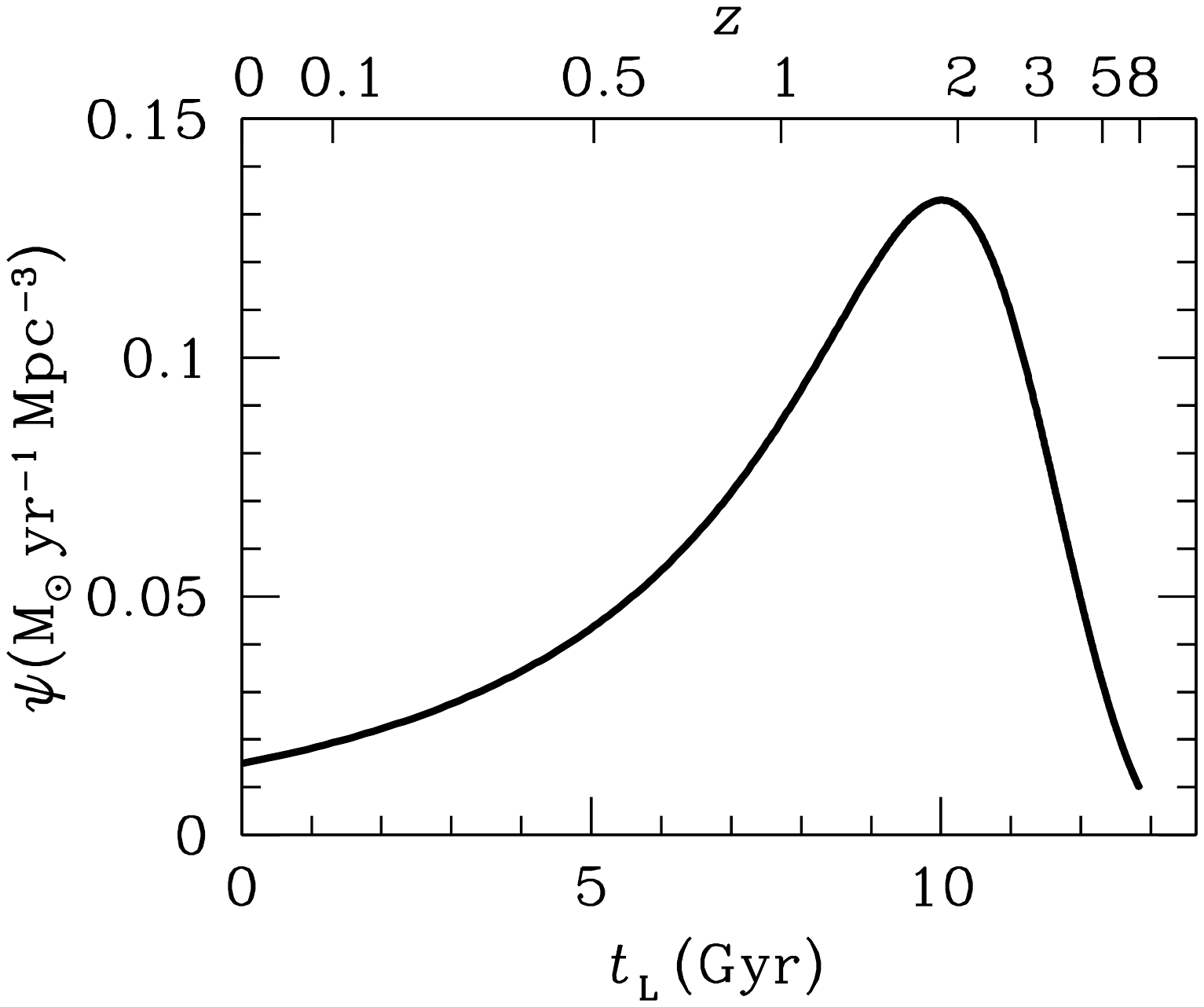} 
\includegraphics[trim = -30 195 0 200,scale=0.45]{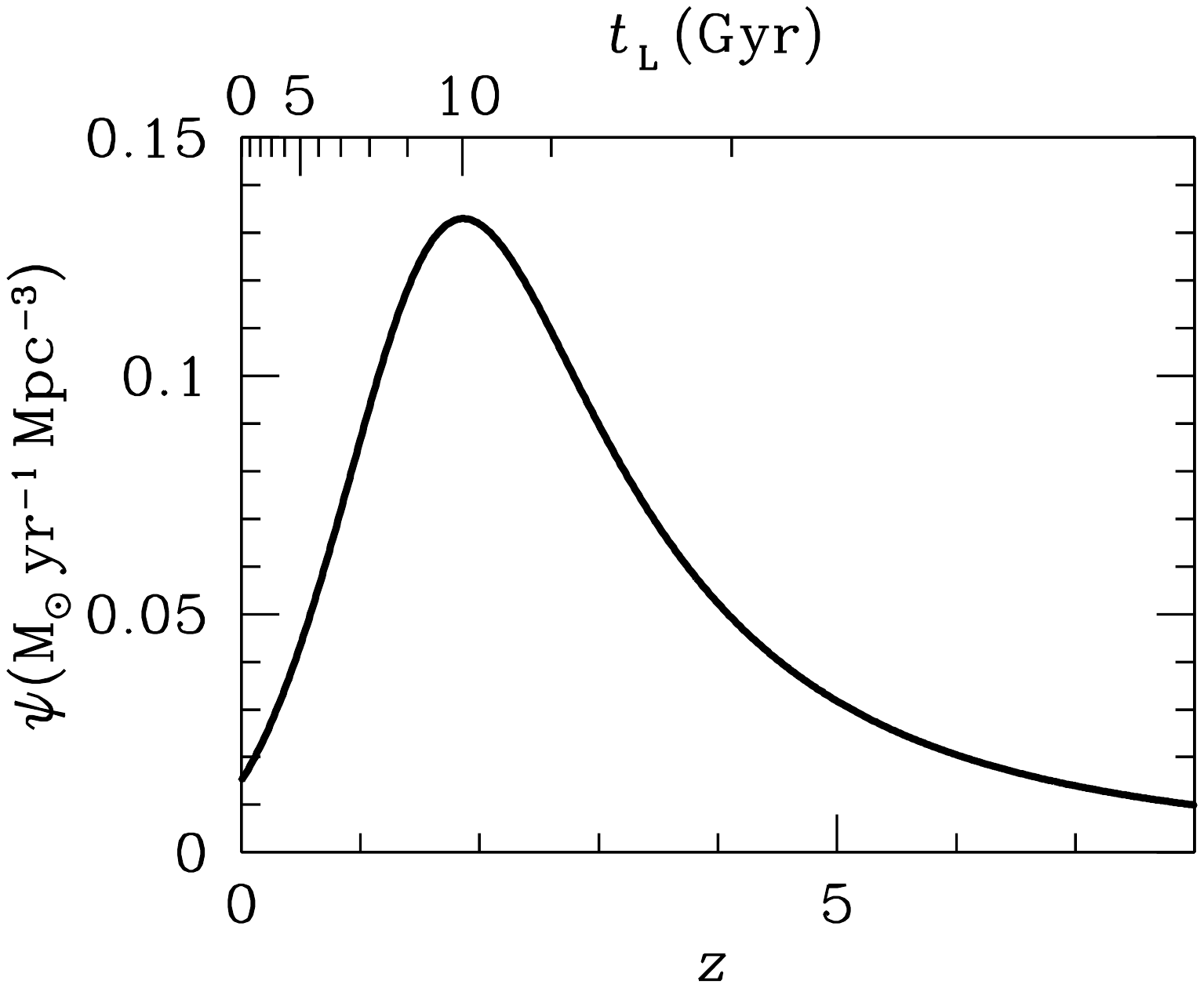}
\includegraphics[trim = -30 160 0 200,scale=0.45]{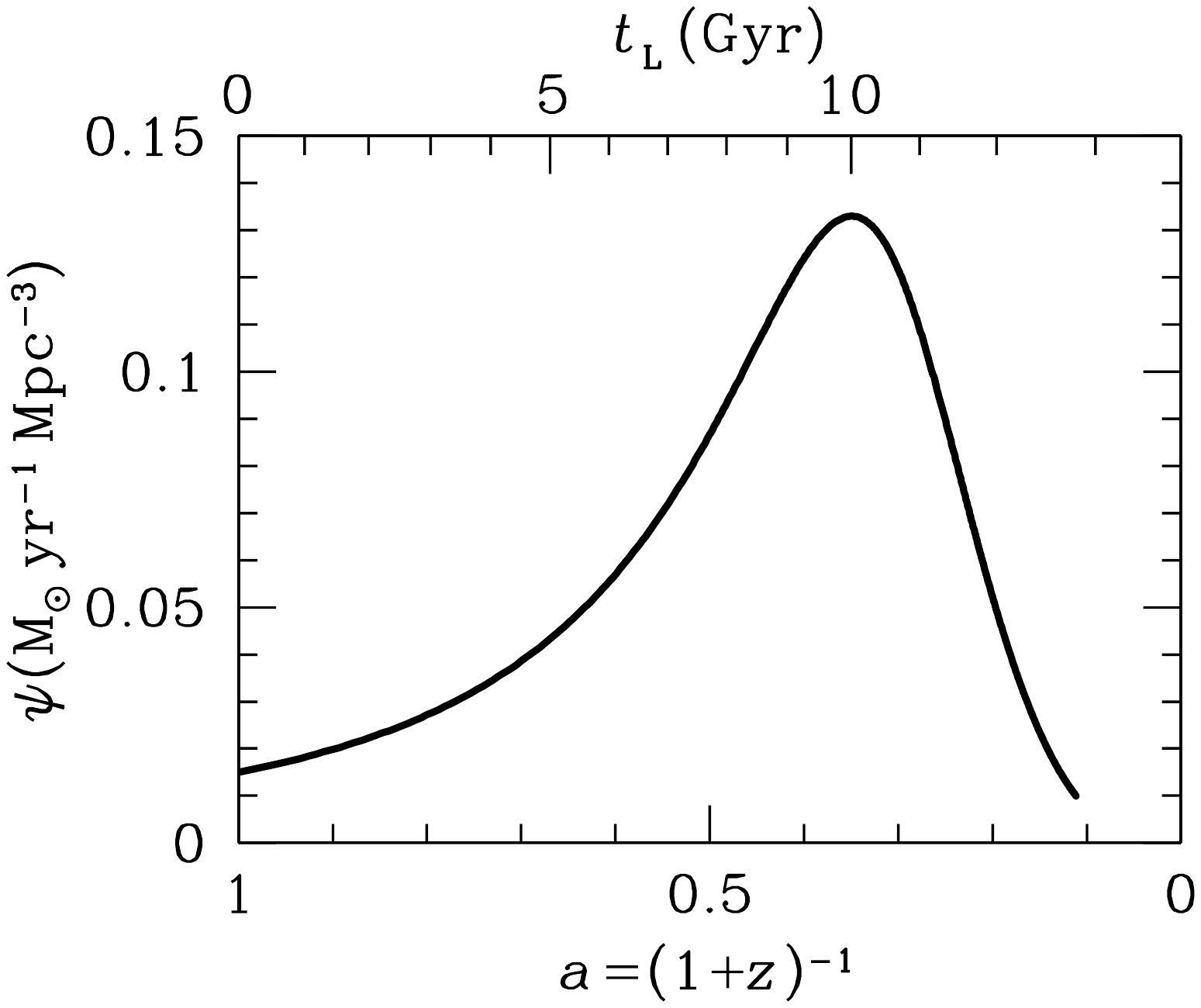} 
\caption{The three panels in this figure show the \citet{mad14} fit to the
  evolving star formation rate density as linear functions of
  lookback time $t_\mathrm{L}$ (top panel), redshift $z$ (middle panel),
  and scale factor $a$ (bottom panel).  Clearly $a$ is a much better
  proxy for $t_\mathrm{L}$ than $z$ is.
\label{fig:sfrd}}
\end{figure}

The observable redshift $z$ is the traditional proxy for lookback time
$t_\mathrm{L}$ in models of cosmological evolution. However, it can be
misleading because the lookback time is an extremely nonlinear
function of redshift when $z \gtrsim 1$.  For example, the top panel
in Figure~\ref{fig:sfrd} shows the \citet{mad14} best fit to the star
formation rate density (SFRD) $\psi$ in solar masses per year of per
(comoving) Mpc$^3$ as a linear of function of lookback time back to $z
= 8$.  This plot accurately displays the fact that only 10\% of
today's stellar mass was assembled before $z \approx 2.9$. The very
nonlinear upper abscissa indicates the redshifts $z$ matching the
lookback times on the lower abscissa.  The time between
$z = 0$ and $z = 1$ is $\approx 7.8 \mathrm{~Gyr}$, but the time
between $z = 2$ and $z = 3$ is only $\approx 1.1 \mathrm{~Gyr}$. The
middle panel plots the same function $\psi$ as a linear function of
redshift, with lookback time now on the very nonlinear upper abscissa.
This plot makes it look like $\gg 10$\% of today's stellar mass was
assembled before $z \approx 2.9$.  The nonlinearity at high redshifts
is primarily caused by the fact that $(1+z)$ is the reciprocal of the
scale factor $a$.  When $\psi$ is plotted as a linear function of $a$
(bottom panel), the upper abscissa showing lookback time is nearly
linear and the plot of $\psi(a)$ looks much more like the plot of
$\psi(t_\mathrm{L})$.

\subsection{Light Travel Distance}

The vacuum speed of light $c$ is invariant, so the light travel
distance $D_\mathrm{T}$ corresponding to lookback time $t_\mathrm{L}$
is
\begin{equation}\label{eqn:pdist}
  \boxed{
    D_\mathrm{T} = c \,t_\mathrm{L} = c \int_0^{z}
      \frac {dz'} { (1+z') H}}~.
\end{equation}
The light travel distance in meters can be interpreted physically as
the number of meter sticks laid end-to-end that the photon must pass
on its journey from the source to the observer.

\begin{figure}[!ht]
\includegraphics[trim = 165 320 0 280,scale=0.8]{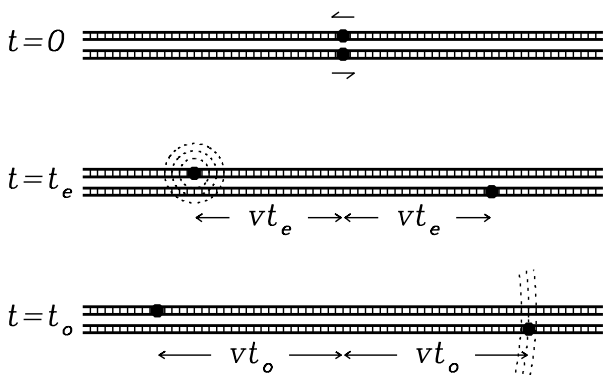} 
\caption{Two trains moving with speed $v$ passed each other at time
  $t = 0$, and the train moving left emitted a sound at time $t_\mathrm{e}$
  when the trains were separated by $d_\mathrm{e} = 2 v t_\mathrm{e}$.
 The sound reached train moving right at time $t_\mathrm{o}$ when the trains
 were separated by $d_\mathrm{o} = 2 v t_\mathrm{o}$. The sound
 travel distance $d_\mathrm{T} = vt_\mathrm{e} + vt_\mathrm{o}$ equals
 neither $d_\mathrm{e}$ nor $d_\mathrm{o}$.
\label{fig:trains}}
\end{figure}

The light travel distance is of limited use in cosmography because it
is the distance between two events occurring at two different proper
times, $t$ and $t_0$.  This limitation can be illustrated by a
nonrelativistic terrestrial example: two trains moving in opposite
directions with speed $v$ passed each other at time $t=0$
(Figure~\ref{fig:trains}).  The horn on one train sounded at time
$t_\mathrm{e}$ when the distance between the trains was $d_\mathrm{e}
= 2 v t_\mathrm{e}$.  The sound reached the other train at a later
time $t_\mathrm{o}$ when the distance between the trains was
$d_\mathrm{o} = 2 v t_\mathrm{o}$.  The sound travel distance is
$d_\mathrm{T} = vt_\mathrm{e} + vt_\mathrm{o} = c_\mathrm{s}
(t_\mathrm{o} -t_\mathrm{e})$, where $c_\mathrm{s} > v$ is the speed
of sound.  Some algebra reveals that $d_\mathrm{T} = d_\mathrm{e}\, [
  c_\mathrm{s} / (c_\mathrm{s} - v)] = d_\mathrm{o} \, [c_\mathrm{s} /
  (c_\mathrm{s} + v)]$ equals neither $d_\mathrm{e}$ nor
$d_\mathrm{o}$.

\subsection{Comoving Coordinates}

Comoving coordinates expand with the universe. The comoving distance
$D_\mathrm{C}$ between any close ($D \ll D_\mathrm{H_0}$) pair of
fundamental observers defined by
\begin{equation}
  D_\mathrm{C} \equiv \frac {D(t)}{a(t)} = D_0~
\end{equation}
(see Equation~\ref{eqn:enlarge}) is independent of $t$ and equals the
present proper distance $D_0$.  Thus comoving rulers are like rubber
bands connecting neighboring fundamental observers, and their markings
agree with rigid rulers today.  At redshift $z$, the vacuum speed of
light in comoving coordinates is $a^{-1}c = (1+z)c$ and
\begin{equation}\label{eqn:ddc}
  d D_\mathrm{C} = (c / a) \, dt = (1+z) c \, d t ~.
\end{equation}

\begin{figure}[!ht]
\includegraphics[trim = 128 270 0 320,scale=0.735]{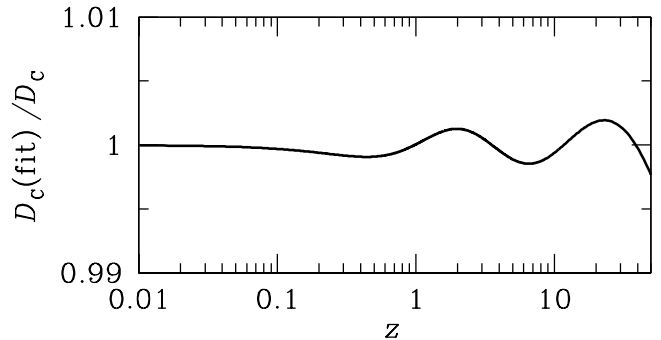} 
\caption{Equation~\ref{eqn:dcfit} fits the comoving distance
 up to $z \approx 50$ within 0.2\% for $\Omega_\mathrm{m} = 0.3$.
 \label{fig:dcfit}}
\end{figure}

In the homogeneous universe, summing over the ``local'' comoving
distances $d D_\mathrm{C}$ yields the global ``line of sight''
comoving distance to a distant source at any redshift $z$:
\begin{align}\label{eqn:dcint}
  D_\mathrm{C} & = c \int_t^{t_0} (1+z) \,dt' \qquad\quad ~~ \nonumber \\
  & = c \int_z^0 (1+z')  \biggl(
  \frac {dt'} {dz'} \biggr) \, dz'~.
\end{align}
Inserting Equations~\ref{eqn:dzdt} and ~\ref{eqn:edef} into
Equation~\ref{eqn:dcint} gives
\begin{equation}\label{eqn:dc}
  \boxed{
D_\mathrm{C} = D_\mathrm{H_0} \int_0^z \frac {dz'}{E(z')}}~.
\end{equation}
Unfortunately, this indefinite integral for $D_\mathrm{C}$ cannot be
expressed in terms of elementary functions, only elliptic integrals
that must be evaluated numerically. It is smooth enough to be
evaluated by Simpson's rule (Appendix~\ref{app:dc}).  Alternatively,
the simple empirical fit
\begin{align}\label{eqn:dcfit}
  D_\mathrm{C}\mathrm{(fit)} \approx &  \,D_\mathrm{H_0} /\,
  [ a/(1-a) + 0.2278 +\nonumber \\
&  \, 0.2070\,(1-a)/(0.785+a) - \nonumber \\
&  \, 0.0158\,(1-a)/(0.312+a)^2]
  ~,
\end{align}
where $a = (1+z)^{-1}$, can be used to avoid the numerical integration
for most astronomical applications.
Equation~\ref{eqn:dcfit} is accurate to within 0.2\% for $z \lesssim
50$ and $\Omega_{0,\mathrm{m}} = 0.3$ (Figure~\ref{fig:dcfit}). 

The comoving distance in meters equals the number of meter sticks laid
end-to-end that would connect the source to the observer today.  It is
the proper distance between us and the source at time $t_0$, so the
comoving distance is the most fundamental  distance for use
in $\Lambda$CDM cosmology.

The ``observable'' universe refers to the sphere in which signals
emitted at any time $t > 0$ and traveling at the vacuum speed of light
could have reached the observer today.  Its light-travel radius is
therefore $ct_0$, and the current comoving radius of the observable
universe is
\begin{equation}
  R_\mathrm{0,C} = D_\mathrm{H_0} \int_0^\infty \frac {dz'} {E(z')}~.
\end{equation}
If $h = 0.7$ and $\Omega_{0,\mathrm{m}} = 0.3$, then $R_\mathrm{0,C}
\approx 3.24\, D_\mathrm{H_0} \approx 3.24 \cdot 2998\, h^{-1}
\mathrm{\,Mpc} \approx 13.9 \mathrm{~Gpc}$.  This number is not
particularly significant, but it is often used to make amusing
calculations like the following: The comoving volume of the observable
universe is $V_0 = 4 \pi R_\mathrm{0,C}^3\, /\, 3 \approx 1.12 \times
10^4 \mathrm{~Gpc}^3 \approx 3.3 \times 10^{80} \mathrm{~m}^3$ and the
present mean density is $\rho_0 \approx 9.2 \times 10^{-27}
\mathrm{~kg~m}^{-3}$, so the total mass of the observable universe is
$M_0 = V_0 \rho_0 \approx 3 \times 10^{54} \mathrm{~kg}$.

\begin{figure}[!ht]
\includegraphics[trim = 130 270 0 255,scale=0.7]{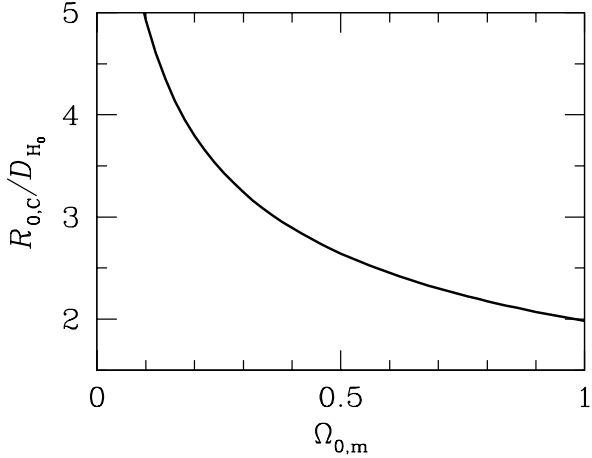} 
\caption{The normalized comoving radius of the observable universe today
  $R_\mathrm{0,C} / D_\mathrm{H_0}$ as a function of $\Omega_{0,\mathrm{m}}$.
\label{fig:rcdh}}
\end{figure}

Figure~\ref{fig:rcdh} shows how the observable radius $R_\mathrm{0,C}$
varies with the normalized matter density $\Omega_{0,\mathrm{m}}$.

The comoving volume $V_\mathrm{C}$ measured in comoving
coordinates is valuable for tracking the cosmic evolution of source
populations because the number of permanent objects (e.g., immortal
fundamental observers, baryons, or galaxies if mergers are ignored) in
a comoving volume element is constant.  (See Appendix~\ref{app:s2ns}
for a sample application of $V_\mathrm{C}$ to calculate source counts from
local luminosity functions.) The Euclidean geometry of
a flat universe implies that the total comoving volume out to redshift
$z$ is 
\begin{equation}\label{eqn:vc}
  \boxed{
    V_\mathrm{C} = \frac{4 \pi D_\mathrm{C}^3}{3} }~, 
\end{equation}
and the comoving volume in a shell covering solid angle $\omega
\mathrm{~sr}$ between $z$ and $z + dz$ is
\begin{equation}
  d V_\mathrm{C} = \omega D_\mathrm{C}^2 \, d D_\mathrm{C} ~.
\end{equation}
Differentiating Equation~\ref{eqn:dc} yields
\begin{equation}
  d D_\mathrm{C} = D_\mathrm{H_0} \,\frac { dz} {E(z)}
\end{equation}
so
\begin{equation}\label{eqn:dvc}
  \boxed{
  d V_\mathrm{C} = \frac {\omega D_\mathrm{C}^2 \,D_\mathrm{H_0}} {E(z)}\, dz }~.
\end{equation}

\subsection{Angular-diameter and Proper-motion Distances}

The flat $\Lambda$CDM universe is Euclidean, so the angular distance
$\theta$ (rad) between two fundamental observers at the same
redshift with comoving transverse separation $l_0$ at comoving
distance $D_\mathrm{C}$ is simply
\begin{equation}
  \theta = \frac{l_0}{D_\mathrm{C}} \quad (\theta \ll 1)~.
\end{equation}
A rigid source (e.g., a transverse meter stick or a gravitationally
bound galaxy) has a fixed proper transverse length $l_\perp$, so its
comoving transverse length is $l_0 =(1+z)\, l_\perp$ and
\begin{equation}\label{eqn:thetadc}
  \theta = \frac {(1+z)l_\perp} {D_\mathrm{C}}~.
\end{equation}

\begin{figure}[!ht]
\includegraphics[trim = 130 270 0 280,scale=0.72]{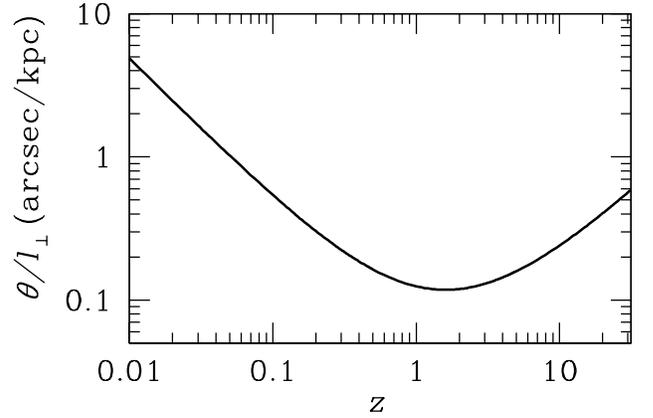}
\caption{The angular diameter $\theta$ of a source with a fixed proper
  diameter $l_\perp = 1 \mathrm{\,kpc}$ has a minimum $\theta \approx
  0\,\farcs118$ at $z \approx 1.6$ in a $\Lambda$CDM universe with $h
  = 0.7$ and $\Omega_\mathrm{m} = 0.3$.
\label{fig:dang}}
\end{figure}

The angular diameter distance $D_\mathrm{A}$  is a ``convenience''
distance defined to satisfy the static Euclidean equation
\begin{equation}\label{eqn:dadef}
D_\mathrm{A} \equiv  \frac{l_\perp}{\theta}
\end{equation}
for a rigid source.
Equations~\ref{eqn:thetadc} and \ref{eqn:dadef} imply
\begin{equation}\label{eqn:dang}
\boxed{
  D_\mathrm{A}  = \frac {D_\mathrm{C}} {(1+z)} }
\end{equation}
in a flat universe. The angular diameter in arcsec ($1
\mathrm{~arcsec} = \pi/ 648000 \mathrm{~rad}$) of a source with fixed
proper diameter $l_\perp = 1 \mathrm{~kpc}$ is shown as a function of
redshift in Figure~\ref{fig:dang}.

The proper motion $\mu$ of a source is its observed angular speed
across the sky:
\begin{equation}
  \mu \equiv \frac{d \theta} {d t}~.
\end{equation}
For example, the radio source in the quasar 3C~279 at $z =
0.5362$ appears to consist of a stationary ``core'' plus moving
components whose proper motions $\mu \sim 0\,\farcs 0005
\mathrm{~yr}^{-1}$ were measured by very long baseline interferometry
\citep{pin03}.

The  proper-motion distance $D_\mathrm{M}$ of a source with
proper transverse velocity $v_\perp = d l_\perp / dt(z)$ defined by
\begin{equation}
    D_\mathrm{M} \equiv \frac {v_\perp} {\mu}
\end{equation}
is also called the ``transverse'' comoving distance. In terms of the
angular diameter distance,
\begin{equation}
  D_\mathrm{M} = \frac {d l_\perp}{d \theta} \frac {dt} { dt(z)}
  = D_\mathrm{A} (1+z) ~,
\end{equation}
where $dt(z)$ is the proper time measured by a fundamental observer at
the source redshift $z$.  The proper-motion distance is just the angular diameter
distance multiplied by the $(1+z)$ time dilation factor.  In a flat
universe,
\begin{equation}\label{eqn:dm}
  \boxed{
    D_\mathrm{M} = D_\mathrm{C}
    }~,
\end{equation}
so the ``line of sight'' comoving distance $D_\mathrm{C}$ equals the
``transverse'' comoving distance $D_\mathrm{M}$ and the two can be
treated simply as a ``the'' comoving distance.  In the non-Euclidean
geometry of a curved universe
$D_\mathrm{M} \neq D_\mathrm{C}$ \citep{hog99}.

\subsection{Luminosities and Fluxes}

Let $L$ be the total (or bolometric) luminosity (emitted power, SI
units W) of an isotropic source measured in the source frame and $F$
be the total flux (power received per unit area, SI units W~m$^{-2}$)
in the observer's frame.  (If the source is not isotropic, $L$ should
be replaced by $4 \pi$ times the power per steradian beamed in the
direction of the observer.)

In a static Euclidean universe, the inverse-square law accounts for
the transverse spatial dilution of photons spread over the area $A_0$
of the spherical surface containing the observer and centered on the
source: $F = L / A_0$.  In the Euclidean but expanding $\Lambda$CDM
universe, the present area of the spherical shell centered on a source
at redshift $z$ and containing the observer is $A_0 = 4 \pi
D_\mathrm{C}^2$.  It is \emph{not} $4 \pi$ times the square of the
distance $D_\mathrm{T}$ covered by the photons in an expanding
universe; the name ``inverse-square (of the distance) law'' is
misleading and ``inverse area law'' would be better.

In an expanding but Euclidean flat universe, the observed flux is
lower than in a static universe because (1) the observed rate at which
photons cross the $t = t_0$ surface centered on the source and
containing the observer is a factor $(1+z)$ lower than the rate at
which they were emitted and (2) the observed energy $E_\mathrm{o} = hc /
\lambda_\mathrm{o}$ of each redshifted photon is a factor $(1+z)$
lower than its energy $E_\mathrm{e} = hc / \lambda_\mathrm{e}$ in the source
frame.  Consequently
\begin{equation}\label{eqn:boloflux}
  F = \biggl(\frac {L} {4 \pi D_\mathrm{C}^2}\biggr)
  \biggl( \frac {1} {1+z} \biggr)^2  ~.
\end{equation}

The luminosity distance $D_\mathrm{L}$ is another ``convenience''
distance defined by the form of the static Euclidean inverse-square
law:
\begin{equation}\label{eqn:dldef}
  \boxed{
   F \equiv \frac {L} {4 \pi D_\mathrm{L}^2} }~.
\end{equation}
Equation~\ref{eqn:boloflux} implies
\begin{equation}\label{eqn:dlum}
  \boxed{
 D_\mathrm{L} = (1+z) D_\mathrm{C}}~.
\end{equation}

\subsection{Comparison of Distance Types}

\begin{figure}[!ht]
\includegraphics[trim = 145 280 0 240,scale=0.75]{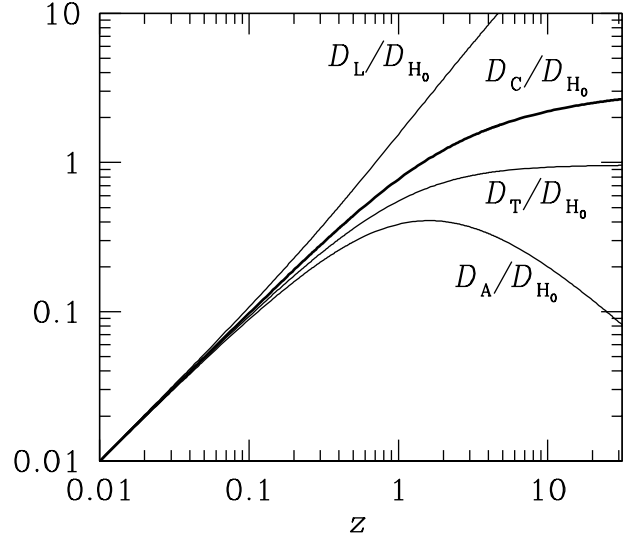} 
\caption{The luminosity distance $D_\mathrm{L}$, the comoving distance
  $D_\mathrm{C}$, the light travel distance $D_\mathrm{T}$, and the
  angular diameter distance $D_\mathrm{A}$, all normalized by the current
  Hubble distance $D_\mathrm{H_0}$, are compared for $\Omega_{0,\mathrm{m}}
  = 0.3$.
\label{fig:dist}}
\end{figure}

Figure~\ref{fig:dist} compares the luminosity distance $D_\mathrm{L}$,
the comoving distance $D_\mathrm{C}$, the light travel distance
$D_\mathrm{T}$, and the angular-diameter distance $D_\mathrm{A}$ in a
$\Lambda$CDM universe with current matter density parameter
$\Omega_{0,\mathrm{m}} = 0.3$.  All of the plotted distances are
normalized by the current Hubble distance $D_\mathrm{H_0} = c / H_0
\approx  2998 \,h^{-1} \mathrm{~Mpc} \approx 4280 \mathrm{~Mpc}$.

\begin{figure}[!ht]
\includegraphics[trim = 145 280 0 330,scale=0.75]{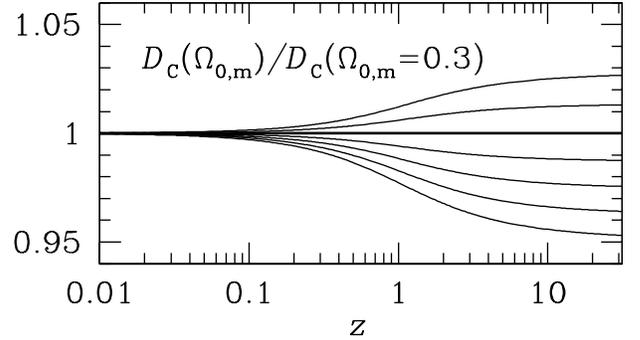} 
\caption{The comoving distances $D_\mathrm{C}$ are plotted
  for $\Omega_{0,\mathrm{m}} = 0.28$ (top curve) through $\Omega_{0,\mathrm{m}}
  = 0.34$ (bottom curve)
 in steps of $\Delta\Omega_{0,\mathrm{m}} = 0.01$,
  all normalized by $D_\mathrm{C}(\Omega_{0,\mathrm{m}} = 0.3)$ (heavy line).
 \label{fig:dcom}}
\end{figure}

These distances also vary slowly and smoothly with
$\Omega_{0,\mathrm{m}}$, but the effect of changing
$\Omega_{0,\mathrm{m}}$ cannot be represented by a scale factor like
$h$.  Figure~\ref{fig:dcom} shows the ratios of
$D_\mathrm{C}(\Omega_{0,\mathrm{m}})$ to
$D_\mathrm{C}(\Omega_{0,\mathrm{m}} = 0.3)$ for matter densities from
$\Omega_{0,\mathrm{m}} = 0.28$ (top curve) through $0.34$ (bottom
curve) that include the best measured $\Omega_{0,\mathrm{m}} = 0.315
\pm 0.013$ \citep{pla16} and its quoted uncertainty.  Near
$\Omega_{0,\mathrm{m}} = 0.3$, $d D_\mathrm{C} / d
\Omega_{0,\mathrm{m}} \approx - 0.006$ at $z=1$ and $d D_\mathrm{C} /
d \Omega_{0,\mathrm{m}} \approx - 0.013$ when $z \gg 1$.

\subsection{Spectral Luminosities and Flux Densities}

The spectral luminosity $L_\nu(\nu)$ of a source is its luminosity per
unit frequency (SI units $\mathrm{W~Hz}^{-1}$).  In this notation, the
subscript $\nu$ just means ``per unit frequency'' in the source frame
and doesn't refer to any particular frequency.  The $\nu$ in
parentheses is the actual frequency in the source frame, so $L_\nu(1.4
\mathrm{~GHz})$ is the power per unit frequency emitted at $\nu = 1.4
\mathrm{~GHz}$ in the source frame.  The spectral flux density $F_\nu(\nu)$
(or $S$) of a source is the observed flux per unit frequency in the
observer's frame (SI units $\mathrm{W~m}^{-2} \mathrm{~Hz}^{-1}$, or
the astronomically practical $\mathrm{Jy} \equiv 10^{-26}
\mathrm{~W~m}^{-2} \mathrm{~Hz}^{-1}$).

All of the photons received in a narrow logarithmic frequency range
centered on frequency $\nu$ were emitted in the equally narrow
logarithmic frequency range centered on $[(1+z)\nu]$, so the
bolometric Equation~\ref{eqn:boloflux} implies
 \begin{equation}\label{eqn:nufnu}
   \nu\, F_\nu(\nu) = [(1+z)\nu] \,
   \frac {L_\nu[(1+z)\nu] } {4 \pi D_\mathrm{L}^2}
 \end{equation}
 and
 \begin{equation}\label{eqn:fnu}
   \boxed{
    F_\nu(\nu) = (1+z)  \frac {L_{\nu}[(1+z)\nu]}
   {4 \pi D_\mathrm{L}^2} }~.
\end{equation}
 The leading factor of $(1+z)$ in Equation~\ref{eqn:fnu} comes from
 bandwidth compression: photons emitted over the frequency range
 $[(1+z)\Delta\nu]$ are squeezed into the frequency range $\Delta\nu$
 in the observer's frame.
 
 The spectral index $\alpha$ between frequencies $\nu_1$ and $\nu_2$
 is defined by
 \begin{equation}\label{eqn:alphadef}
   \alpha(\nu_1,\nu_2) \equiv + \frac {\ln[L(\nu_1)/L(\nu_2)]}
         {\ln(\nu_1/\nu_2)}~.
 \end{equation}
 [Beware that some authors define $\alpha$ with the opposite sign.]
 In terms of $\alpha[\nu,(1+z)\nu]$, Equation~\ref{eqn:fnu}
 becomes
 \begin{equation}\label{eqn:fluxdensity}
     F_\nu(\nu) = (1+z)^{\alpha+1}\, \frac {L_\nu(\nu)}
     { 4 \pi D_\mathrm{L}^2} ~.
   \end{equation}

\begin{figure}[!ht]
\includegraphics[trim = 125 270 0 240,scale=0.7]{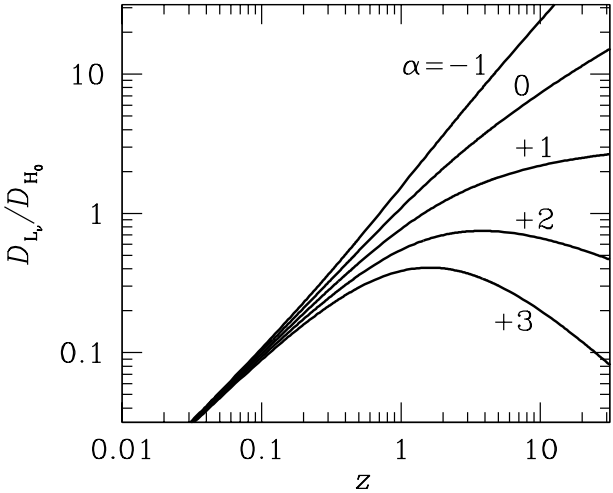} 
\caption{Spectral luminosity distances normalized by the current
  Hubble distance are plotted for spectral indices $\alpha = -1$, $0$,
  $+1$, $+2$, and $+3$.  $\Omega_{0,\mathrm{m}} = 0.3$ in all cases.
\label{fig:dlnu}}
\end{figure}

 If the spectral luminosity distance $D_{\mathrm{L}_\nu}$ is
 defined by analogy with Equation~\ref{eqn:dldef}:
 \begin{equation}\label{eqn:dlnudef}
   F_\nu \equiv
   \frac {L_\nu} {4 \pi D_{\mathrm{L}_\nu}^2}, 
 \end{equation}
 then Equation~\ref{eqn:fnu} implies
 \begin{equation}
   D_{\mathrm{L}_\nu} = D_\mathrm{L} \, (1+z)^{-(\alpha+1)/2}~.
 \end{equation}
 Figure~\ref{fig:dlnu} shows $D_{\mathrm{L}_\nu}(z) / D_\mathrm{H_0}$
 for a range of spectral indices.  Steep-spectrum ($\alpha \approx
 -1$) synchrotron sources have $D_{\mathrm{L}_\nu} \approx
 D_\mathrm{L}$ and consequently are quite faint at high redshifts.
 Flat-spectrum self-absorbed synchrotron sources and optically thin
 free-free emitters ($\alpha \approx 0$) are only slightly stronger.
 For a source with $\alpha = +1$, $D_{\mathrm{L}_\nu} = D_\mathrm{C}$.
 Blackbody emission in the long wavelength Rayleigh-Jeans limit has
 $\alpha \approx +2$. Optically thin dusty galaxies have $\alpha
 \sim +3$ at wavelengths $0.1 \lesssim \lambda_\mathrm{e} \lesssim 1
 \mathrm{~mm}$, so their $D_{\mathrm{L}_\nu} \sim
 D_\mathrm{A}$ and their submillimeter flux densities are nearly
 independent of redshift over a broad range centered on $z \sim
 1.6$.

 The spectral flux density $F_\lambda$ measured per unit wavelength
 (SI units $\mathrm{W~m}^{-3}$) is related to $F_\nu$ by
 \begin{equation}
   \vert F_\lambda \,d \lambda \vert = \vert F_\nu \,d \nu \vert
 \end{equation} 
 so
 \begin{equation}\label{eqn:fnuflambda}
   F_\lambda = \frac {c}{\lambda^2} \,F_\nu \mathrm{~~and~~}
   L_\lambda = \frac {c} {\lambda^2} \, L_\nu. 
 \end{equation}
 The wavelength counterparts 
 of Equations~\ref{eqn:nufnu} and \ref{eqn:fnu} are
 \begin{equation}
   \lambda \, F_\lambda(\lambda) =
   \frac {[\lambda/(1+z)]\, L_\lambda[\lambda/(1+z)]}
   {4 \pi D_\mathrm{L}^2}~.
 \end{equation}
 and
 \begin{equation}\label{eqn:flambda}
\boxed{   F_\lambda(\lambda) = (1+z)^{-1}\,
   \frac { L_\lambda [\lambda/(1+z)]} {4 \pi D_\mathrm{L}^2} } ~.
 \end{equation}

 \subsection{Magnitudes and $K$ Corrections}

\begin{figure}[!ht]
\includegraphics[trim = 125 270 0 235,scale=0.7]{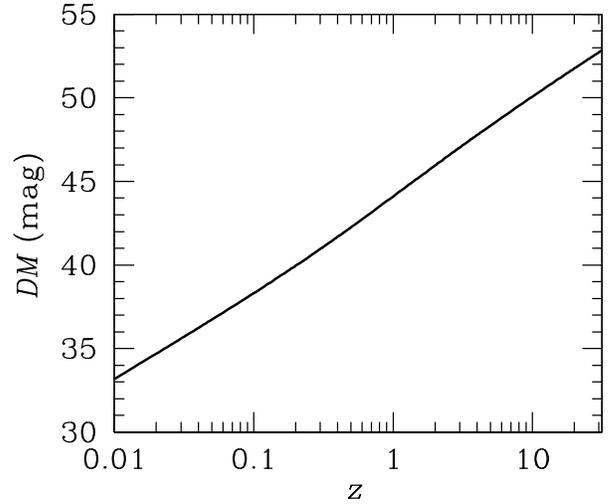} 
\caption{The bolometric distance modulus $DM$ as a function of redshift $z$
  for $h=0.7$, $\Omega_{0,\mathrm{m}} = 0.3$.
\label{fig:distmod}}
\end{figure}

 The apparent magnitude $m$ and absolute magnitude $M$ of a source
 are related by
 \begin{equation}\label{eqn:magdef}
\boxed{
   m - M = 5 \log_{10}\biggl(
   \frac{D_\mathrm{L}}{10 \mathrm{~pc}} \biggr) + K}~,
\end{equation}
 where
 \begin{equation}
   DM \equiv  5 \log_{10}\biggl(
   \frac{D_\mathrm{L}}{10 \mathrm{~pc}} \biggr)
 \end{equation}
 is the bolometric distance modulus (Figure~\ref{fig:distmod}) in
 magnitudes ($1 \mathrm{~mag} \equiv 10^{-0.4}$). The $K$ correction
 converts the apparent magnitude measured through a filter
 covering a fixed wavelength range \citep{oke68,hog02} in the observer's
 frame to yield the absolute magnitude over the same wavelength range
 in the source frame.  The bolometric $K$ correction is zero.  There
 are many magnitude systems covering different bandpasses and having
 different zero points \citep{bla07} relating $m=0$ to flux density.

 Equation~\ref{eqn:fnu} implies that 
 \begin{eqnarray}
   K =  -2.5 \log_{10} \biggl\{(1+z)
   \frac {L_\nu[(1+z)\nu]} {4 \pi D_\mathrm{L}^2} \biggr\} + \nonumber \\
    2.5 \log_{10} \biggl\{
   \frac {L_\nu(\nu)} {4 \pi D_\mathrm{L}^2} \biggr\} \qquad\qquad\qquad\quad \nonumber \\
\boxed{
K  =  -2.5 \log_{10} \biggl\{ (1+z)
   \frac {L_\nu[(1+z)\nu]} {L_\nu(\nu)} \biggr\} }~.
 \end{eqnarray}
 In terms of the spectral index $\alpha$, 
 \begin{align}\label{eqn:kalpha}
   K & = -2.5 \log_{10} \bigl[(1+z)^{\alpha+1}\bigr] \quad ~~\nonumber \\
  & = -2.5 \,(\alpha+1) \log_{10}(1+z)~.
 \end{align}
In terms of wavelengths, 
 \begin{equation}
   \boxed{
     K = -2.5 \log_{10} \biggl\{ (1+z)^{-1}
 \frac {L_\lambda[\lambda/(1+z)]} {L_\lambda(\lambda)} \biggr\} }~.
 \end{equation}

 \subsection{Spectral Lines}

Spectral lines are narrow (line width $\Delta \nu \ll \nu$) emission or
absorption features in the spectra of gaseous or ionized sources.  The
total line luminosity $L$ is related to the total line flux $F$ and
the line flux density $F_\nu$ by Equation~\ref{eqn:dldef}, so
\begin{equation}\label{eqn:linelum}
  L = 4 \pi D_\mathrm{L}^2 F = 4 \pi D_\mathrm {L}^2 \, F_\nu \Delta \nu~.
\end{equation}
In Equation~\ref{eqn:linelum}, the SI units of $F$ are
$\mathrm{W~m}^{-2} = 10^{26} \mathrm{~Jy~Hz}$.  However, line
fluxes are often reported in the dimensionally confusing units
$\mathrm{Jy~km~s}^{-1}$ based on the nonrelativistic Doppler
equation
\begin{equation}\label{eqn:Doppler}
  \frac {\Delta \nu} {\nu} \approx \frac {\Delta v} {c} \ll 1~.
\end{equation}
Solving Equation~\ref{eqn:Doppler} for $\Delta v$ yields the 
factor needed to convert from $\mathrm{Jy~km~s}^{-1}$ to Jy~Hz:
\begin{equation}
  1 \mathrm{~km~s}^{-1} \approx \frac {\nu} {299792} \mathrm{~Hz}~.
  \end{equation}
 See \citet{car13} for a detailed discussion of this equation and its uses.

 \subsection{Total Intensity and Specific Intensity}

The total intensity or bolometric brightness $B$ of a source is the
power it emits per unit area per unit solid angle $\omega$ (SI units
$\mathrm{W~m}^{-2} \mathrm{~sr}^{-1}$).  In a \emph{static} Euclidian
universe, intensity is conserved along any ray passing through empty
space and the brightness $B_0$ seen by an observer at rest with
respect to the source equals $B$.  For a source at redshift $z$ in the
Euclidean but expanding $\Lambda$CDM universe, the observed brightness
$B_0$ can be calculated with the aid of Equations~\ref{eqn:dang} and
\ref{eqn:dlum}.  For a source at any comoving distance $D_\mathrm{C}$,
expansion multiplies its luminosity distance by $(1+z)$ and divides
its angular diameter distance by $(1+z)$, so
\begin{equation}
  \frac {B_0} {B} = \frac {F_0} {F}
  \frac {\omega} {\omega_0} = (1+z)^{-2} (1+z)^{-2}
\end{equation}
and
\begin{equation}\label{eqn:brt}
  \boxed{
    \frac {B_0}{B} = (1+z)^{-4}
    }~.
\end{equation}

The specific intensity or spectral brightness $B_\nu(\nu)$ of a source
is its power per unit frequency per unit solid angle (SI units
$\mathrm{W~m}^{-2} \mathrm{~Hz}^{-1} \mathrm{~sr}^{-1}$) at frequency
$\nu$ in the source frame.

All of the photons received in a narrow logarithmic frequency range
centered on $\nu$ were emitted in the equally narrow logarithmic
frequency range centered on $[(1+z)\nu]$, so the bolometric brightness
Equation~\ref{eqn:brt} implies
 \begin{equation}\label{eqn:sb}
 \nu B_{\nu0} =  \frac {[(1+z)\nu]\,B_\nu[(1+z)\nu]} {(1+z)^4}~.
 \end{equation}
 \begin{equation}
\boxed{
   B_{\nu0} (\nu) = (1+z)^{-3}B_\nu[(1+z)\nu] } ~.
 \end{equation}
 For a source with spectral index $\alpha$,
 \begin{equation}\label{eqn:specbrt}
  B_{\nu0} (\nu) = (1+z)^{\alpha - 3} B_\nu(\nu) ~.
\end{equation}

The Planck brightness spectrum of a blackbody source at temperature $T$ is
\begin{equation}
  \mskip-2mu  B_\nu(\nu \,\vert\, T) =
  \frac {2 h \nu^3} {c^2} \Biggl[ \exp \biggl(
    \frac{h \nu} {kT} \biggr) -1 \Biggr]^{-1} .
\end{equation}
If the source is at redshift $z$, the observed spectrum
\begin{align}
  B_{\nu0}(\nu) & = (1+z)^{-3} B_\nu[(1+z)\nu \,\vert\, T] \qquad\qquad \nonumber \\
  & = \frac {2 h } {c^2}
  \biggl[ \frac {(1+z) \nu} {(1+z)} \biggr]^3 
  \biggl\{ \exp \biggl[ \frac {h \nu} {k}
    \frac {(1+z)} {T} \biggr] -1 \biggr\}^{-1}
\end{align}
is just the Planck spectrum  $B_\nu(\nu \vert T_0)$ of a blackbody
with temperature $T_0 = T / (1+z)$.  The source of the
$T_0 \approx 2.73 \mathrm{~K}$ CMB seen today is the $T \sim 3000
\mathrm{~K}$ blackbody surface of last scattering when the universe
became transparent at $z \sim 1100$.
 
\subsection{Nonrelativistic Approximation Errors}

\begin{figure}[!ht]
\includegraphics[trim = 135 280 0 230,scale=0.70]{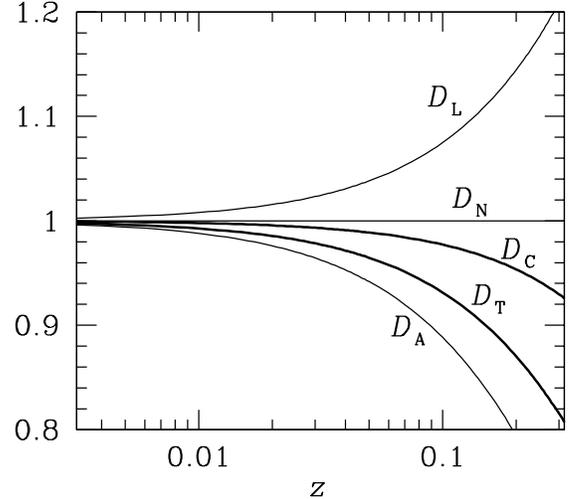} 
\caption{The ratios of the relativistically correct distances
  $D_\mathrm{L}$, $D_\mathrm{C}$, $D_\mathrm{T}$, and $D_\mathrm{A}$
  to the nonrelativistic distance approximation $D_\mathrm{N} \equiv c
  z / H_0$ at low redshifts $z$ indicate the errors that can result
  from using $D_\mathrm{N}$.
\label{fig:drat}}
\end{figure}

In the low-redshift limit, it is tempting to calculate intrinsic
source parameters from the observables using the nonrelativistic
distance approximation
\begin{equation}
  D_\mathrm{N} \equiv \frac {cz}{H_0}~.
\end{equation}
Figure~\ref{fig:drat} displays ratios of the relativistically correct
luminosity distance $D_\mathrm{L}$, comoving distance $D_\mathrm{C}$,
light travel distance $D_\mathrm{T}$, and angular diameter distance
$D_\mathrm{A}$ in a $\Lambda$CDM universe with $\Omega_{0,\mathrm{m}} =
0.3$ to $D_\mathrm{N}$. 

For example, the bolometric luminosity $L$ of a source at redshift $z$
is proportional to $D_\mathrm{L}^2$, so the luminosity calculated
using $D_\mathrm{N}$ instead will be too low by a factor of
$(D_\mathrm{L} / D_\mathrm{N})^2$.  It will be 5\% too low for a
source at redshift $z \approx 0.032$ ($cz \sim 10^4
\mathrm{~km~s}^{-1}$), and the maximum redshift at which the
luminosity error is $<10$\% is $z \approx 0.065$ ($cz \sim 2\times
10^4 \mathrm{~km~s}^{-1}$).

The calculated linear size of a source is proportional to
$D_\mathrm{A} < D_\mathrm{N}$, so using $D_\mathrm{N}$ will
overestimate source size by 5\% at $z \approx 0.043$ ($cz \sim 1.3
\times 10^4 \mathrm{~km~s}^{-1}$) and 10\% at $z \approx 0.089$ ($cz
\sim 2.7 \times 10^4 \mathrm{~km~s}^{-1}$).

Such errors are systematic, so they can easily dominate the Poisson
errors in statistical properties of large source
populations. Luminosity functions are particularly vulnerable because
they depend on the maximum redshifts at which sources \emph{could}
have remained in the flux-limited population, not just the actual
source redshifts.

\subsection{Example Calculation: A Single Source}

\begin{figure}[!ht]
\includegraphics[trim = 40 160 0 130,scale=0.45]{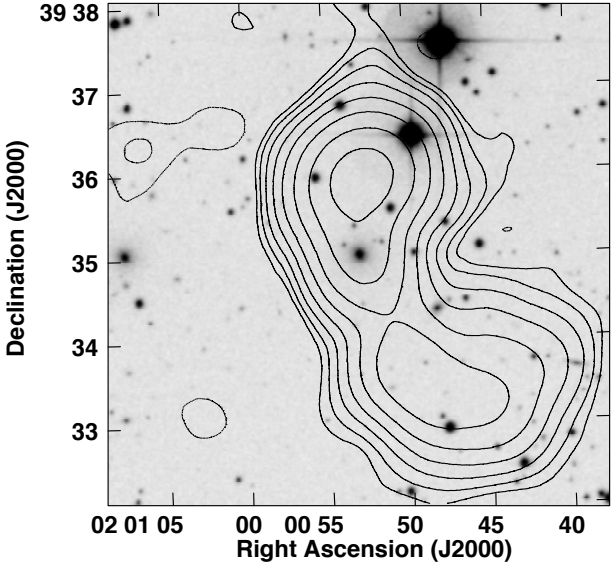} 
\caption{The bent radio source 4C\,+39.05 (1.4~GHz contours at
  $\pm 1,\,2,\,4,\dots,\,128 \mathrm{~mJy~beam}^{-1}$ in a $45''$ FWHM
  Gaussian beam) originated in
  the elliptical galaxy 2MASX 02005301+3935003 at the center of this
  $6' \times 6'$ optical finding chart (gray scale).
\label{fig:4C39.05}}
\end{figure}

The bent triple radio source 4C\,+39.05 (Figure~\ref{fig:4C39.05}) is
identified with the galaxy 2MASX 02005301+3935003 at $z \approx 0.0718$.
Its 1.4~GHz flux density is $S \approx 635 \mathrm{~mJy}$ and its
angular diameter is $\theta \approx 200''$.

The light-travel time from 4C\,+39.05 can be found by inserting
$z = 0.0718$ and $\Omega_{0,\Lambda} \approx 1 - \Omega_{0,\mathrm{m}}
= 1 - 0.3 = 0.7$ into Equation~\ref{eqn:tlapprox}; it is $t_\mathrm{L}
/ t_{\mathrm{H}_0} \approx 0.0682$. For $h = 0.7$, $t_{\mathrm{H}_0}
\approx 14 \mathrm{~Gyr}$ (Equation~\ref{eqn:th0}) and $t_\mathrm{L}
\approx 9.5 \times 10^8 \mathrm{~yr}$.

For $h = 0.7$ and $\Omega_{0,\mathrm{m}} = 0.3$, its comoving distance
from either Equation~\ref{eqn:dc} or Equation~\ref{eqn:dcfit} is
$D_\mathrm{C} \approx 303 \mathrm{~Mpc}$.  The comoving volume within
the sphere ($\omega = 4 \pi \mathrm{~sr}$) of this radius is
(Equation~\ref{eqn:vc}) $V_\mathrm{C} = 4 \pi D_\mathrm{C}^3 / 3
\approx 4 \cdot 3.14 \cdot (303 \mathrm{Mpc})^3 / 3 \approx 1.17
\times 10^8 \mathrm{~Mpc}^3$.

The angular-size distance (Equation~\ref{eqn:dang}) to 4C\,+39.05 is
$D_\mathrm{A} = D_\mathrm{C} / (1+z) \approx 283 \mathrm{~Mpc}$, so
its projected linear size is $l_\perp = \theta D_\mathrm{A} \approx
200 \mathrm{~arcsec} \cdot \pi / 648000 \mathrm{~rad/arcsec} \cdot 283
\mathrm{~Mpc} \approx 0.27 \mathrm{~Mpc}$.  Notice that the angular
diameter of this source would remain $\theta > 30''$ if it were
moved to \emph{any} redshift (Figure~\ref{fig:dang}).

At 4.85~GHz 4C\,+39.05 has  flux density $S \approx 200 \mathrm{~mJy}$,
so its spectral index (Equation~\ref{eqn:alphadef}) is
\begin{equation}
  \alpha = + \frac{\ln (635 \mathrm{~mJy} / 200 \mathrm{~mJy})}
         {\ln (1.4 \mathrm{~GHz} / 4.85 \mathrm{~GHz})} \approx -0.9~.
\end{equation}
The absolute spectral luminosity of 4C\,+39.05 at $\nu = 1.4
\mathrm{~GHz}$ in the source frame can be obtained by solving
Equation~\ref{eqn:fluxdensity} for $L_\nu(\nu)$:
\begin{equation}
  L_\nu(\mathrm{1.4~GHz}) = 4 \pi D_\mathrm{L}^2 (1+z)^{-\alpha -1}
  F_\nu(\mathrm{1.4~GHz})~,
\end{equation}
where $D_\mathrm{L} = (1+z) D_\mathrm{C} \approx 1.0718 \cdot 303
\mathrm{~Mpc} \approx 325 \mathrm{~Mpc} \cdot 3.0857\times10^{22}
\mathrm{~m~Mpc}^{-1} \approx 1.00 \times 10^{25} \mathrm{~m}$
(Equation~\ref{eqn:dlum}).
\begin{eqnarray}
  L_\nu(\mathrm{1.4~GHz}) \approx 4 \cdot 3.14 \cdot
  (1.00 \times 10^{25} \mathrm{~m})^2 \cdot \qquad\qquad~ \nonumber \\
  \quad 1.0718^{-0.1} \cdot 635 \mathrm{~mJy} \cdot 10^{-29} \mathrm{\,W\,m}^{-2}
  \mathrm{\,Hz}^{-1} \mathrm{\,mJy}^{-1} ~~~\nonumber \\
  L_\nu\mathrm{(1.4~GHz)} \approx 7.9 \times 10^{24} \mathrm{\,W\,Hz}^{-1}~.
  \qquad \qquad\qquad ~~~~
\end{eqnarray}

The $\lambda \approx 2.2 ~\mu\mathrm{m}$ apparent magnitude of the host galaxy
is $k_\mathrm{20fe} \approx 11.788$.  The $K$ correction at this
wavelength is $K \approx -6.0 \log_{10}(1+z)$ independent of galaxy
type and is valid for any $z \lesssim 0.25$ \citep{koc01}.  At $z =
0.0718$, $K \approx -0.181$ and Equation~\ref{eqn:magdef} can be used
to calculate the absolute magnitude of the host galaxy at $\lambda
\approx 2.2 ~\mu\mathrm{m}$ in the source frame:
\begin{eqnarray}
  K_\mathrm{20fe} \approx k_\mathrm{20fe}  - 5 \log_{10}
  \biggl( \frac {D_\mathrm{L}}{10 \mathrm{~pc}} \biggr) -K \qquad \nonumber \\
  \approx 11.778 - 5 \log_{10} (303\times10^6 \mathrm{~pc} / 10 \mathrm{~pc})
  + 0.181 \nonumber \\
  \approx 11.778 -37.407 + 0.181 \approx -25.448~. \qquad\quad~~
\end{eqnarray}
The $k_\mathrm{20fe}=0$ flux density is $S = 666.7 \pm 12.6
\mathrm{~Jy}$ at $\nu_\mathrm{o} \approx 1.390 \times 10^{14}
\mathrm{~Hz}$ ({\tt http://www.ipac.caltech.edu/\allowbreak
  2mass/releases/\allowbreak allsky/faq.html\#jansky}) so
$K_\mathrm{20fe} = 0$ corresponds to a spectral luminosity
\begin{align}
  L_\nu \approx & \, 4 \pi \,(10 \mathrm{pc} \,\cdot\, 3.0857 \times 10^{16}
  \mathrm{~m~pc}^{-1})^2\, \cdot \nonumber \\
  & \, 666.7 \mathrm{~Jy} \,\cdot\, 10^{-26}
  \mathrm{~W~m}^{-2} \mathrm{~Hz}^{-1} \mathrm{~Jy}^{-1} \nonumber \\
  \approx & \,
  7.98 \times 10^{12} \mathrm{~W~Hz}^{-1} 
  \end{align}
and $K_{\mathrm 20fe} = -25.448$ corresponds to a spectral luminosity
\begin{align}
  L_\nu & \approx 7.98 \times 10^{12} \mathrm{~W~Hz}^{-1} \,\cdot\,
  10^{0.4 \,\cdot \,25.448} \nonumber \\
  & \approx 1.2 \times 10^{23} \mathrm{~W~Hz}^{-1} \qquad\qquad\qquad~
\end{align}
at $\nu_\mathrm{e} \approx 1.390 \times 10^{14} \mathrm{~Hz}$ in the
source frame.

From Equation~\ref{eqn:specbrt}, the observed 1.4~GHz spectral
brightness of the $\alpha \approx -0.9$ radio source at $z \approx
0.0718$ is lower than its 1.4~GHz specific intensity in the source
frame by the factor
\begin{equation}
  \frac {B_{\nu0}} {B_\nu} = (1+z)^{\alpha-3}
  \approx 1.0718^{-3.9} \approx 0.76~.
\end{equation}

\subsection{Example Calculation: A Source Population}

The 1.4~GHz spectral luminosity of a star-forming galaxy is a linear
and dust-unbiased tracer of the recent star formation rate (SFR)
\citep{mur11}:
\begin{equation}
  \Biggl( \frac {\mathrm{SFR}} {M_\odot\,\mathrm{yr}^{-1}} \Biggr) =
  1.0 \pm 0.1 \times 10^{-21} \Biggl( \frac {L_\mathrm{1.4\,GHz}}
  {\mathrm{W\,Hz}^{-1}} \Biggr)~.
\end{equation}
If the comoving space density of 1.4~GHz sources in star-forming
galaxies is $\rho(L_\mathrm{1.4\,GHz})$, then the local
luminosity-weighted spectral power density function
(Equation~\ref{eqn:udex}) is
\begin{equation}
    U_\mathrm{dex}(L_\nu \,\vert\, z) 
  = \ln(10)  L_\nu^2  \rho(L_\nu \,\vert\, z)~.
\end{equation}

\begin{figure}[!ht]
\includegraphics[trim = 55 270 0 270,scale=0.75]{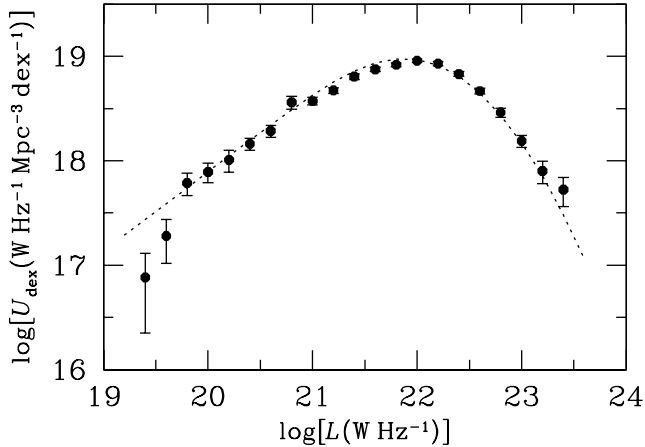} 
\caption{The local 1.4~GHz spectral energy density function
  $U_\mathrm{dex}$ of galaxies whose radio emission is powered
  primarily by star formation, not by AGNs. 
\label{fig:logurads}}
\end{figure}

The observed $U_\mathrm{dex}(L_\mathrm{1.4\,GHz}\, \vert\, z \approx
0)$ \citep{con18} is shown by the data points in
Figure~\ref{fig:logurads}, and it can be approximated by the function
(dotted curve in Figure~\ref{fig:logurads})
\begin{align}
  U_\mathrm{dex}(L_\mathrm{1.4\,GHz}\, \vert \,z \approx 0) \approx &
   ~L_\mathrm{1.4~GHz} (\mathrm{W\,Hz}^{-1})~  \cdot \nonumber \\
  4.0 \times 10^{-3} \mathrm{~Mpc}^{-3} & \mathrm{~dex}^{-1} 
  ~\biggl( \frac {L_\mathrm{1.4\,GHz}} { L_\nu^*} \biggr)^\beta \cdot\nonumber \\
  \exp \Biggl[ - \frac{1}{2 \sigma^2} \log^2
    \Biggl( 1 & +  \frac {L_\mathrm{1.4\,GHz}} {L_\nu^*} \Biggr) \Biggr]~,
\end{align}
where $L_\mathrm{1.4\,GHz}^* = 1.7 \times 10^{21} \mathrm{~W~Hz}^{-1}$,
$\beta = -0.24$, and $\sigma = 0.585$.

\begin{figure}[!ht]
\includegraphics[trim = 50 270 0 290,scale=0.75]{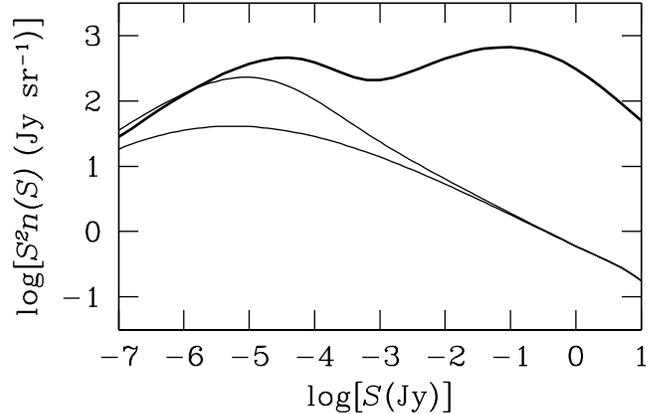} 
\caption{The brightness-weighted count $S^2 n(S)$ of all 1.4~GHz radio
  sources is shown by the heavy curve.  The lower light curve is the
  count of radio sources that would be produced by a non-evolving
  population of star-forming galaxies, and the upper light curve is
  the count that would result from pure luminosity evolution
  consistent with the \citet{mad14} formula for the evolving SFRD
  $\psi(z)$.
\label{fig:s2ns}}
\end{figure}

The evolution of the SFRD $\psi(z)$ can be constrained by
comparing the observed brightness-weighted 1.4~GHz source count $S^2
n(S)$ \citep{con12} shown as the heavy curve in Figure~\ref{fig:s2ns}
with counts predicted by Equation~\ref{eqn:s2ns} for various evolving
$U_\mathrm{dex} (L_\mathrm{1.4\,GHz} \,\vert \,z)$.  The total count
has two peaks, the peak at $\log[S(\mathrm{Jy})] \sim -1$ produced by
AGN-powered radio sources and the peak at $\log[S(\mathrm{Jy})] \sim
-4.5$ attributed to star-forming galaxies.

For the case of no evolution in the SFRD $\psi$, inserting
$U_\mathrm{dex}(L_\mathrm{1.4\,GHz} \,\vert \,z) =
U_\mathrm{dex}(L_\mathrm{1.4\,GHz} \,\vert\, z \approx 0)$ and the median
spectral index $\alpha \approx -0.7$ yields the lower light curve in
Figure~\ref{fig:s2ns}.  The slope of this curve is $\approx -0.5$ at
high flux densities because the stronger star-forming galaxies are at
such low redshifts that their counts approach the static Euclidean
limit $S^{5/2}n(S) = \mathrm{\,constant}$.

The \citet{mad14} model for the evolution of the SFRD $\psi$ is shown
in Figure~\ref{fig:sfrd}.  Their result might be interpreted in terms of
pure luminosity evolution: the comoving density of star-forming
galaxies is constant and the luminosity of each galaxy is proportional
to $\psi$, so $U_\mathrm{dex}(L_\mathrm{1.4\,GHz} \,\vert\, z) =
[\psi(z) / \psi(0)] \, U_\mathrm{dex}(L_\mathrm{1.4\,GHz} \,\vert\, z \approx
0)$.  Inserting this $U_\mathrm{dex}(L_\mathrm{1.4\,GHz} \,\vert\, z)$
into Equation~\ref{eqn:s2ns} yields the higher thin curve in
Figure~\ref{fig:s2ns}.  Both thin curves agree for the low-redshift
sources at high flux densities, but the evolving SFRD produces a peak
at $\log[S(\mathrm{Jy})] \sim -5$ that is much closer to the observed
faint-source count.

\acknowledgments

We thank Ken Kellermann, Eric Murphy, Kristina Nyland, and Mark
Whittle for their valuable comments and suggestions.

\clearpage


\appendix

\section{Analytic Approximations for Lookback Time and Age}\label{app:tlkbk}

Equation~\ref{eqn:tl} for the lookback time $t_\mathrm{L}$:
\begin{equation}
  t_\mathrm{L} = \frac{D_\mathrm{L}}{c} = t_{\mathrm{H}_0} \int_0^z \frac {dz'}{(1+z')E(z')}
\end{equation}
can be integrated analytically if the currently small radiation term
$\Omega_{0,\mathrm{r}}$ is ignored in $E(z')$.  Then
\begin{equation}\label{eqn:aeqn}
\frac {t_\mathrm{L}} {t_{\mathrm{H}_0}} \approx
\int_0^z \frac {dz'}
{(1 + z') [\Omega_{0,\mathrm{m}}(1 + z')^3 + \Omega_{0,\Lambda}]^{1/2}}~.
\end{equation}
Substituting $\Omega_{0,\Lambda} = 1 -
\Omega_{0,\mathrm{m}}$ and $x = (1+z')^{-3/2}$ reduces
Equation~\ref{eqn:aeqn} to 
\begin{equation}
\frac {t_\mathrm{L}} {t_{\mathrm{H}_0}} \approx \frac {2}{3\,\Omega_{0,\Lambda}^{1/2}}
\int_{(1+z)^{-3/2}}^1 \frac{dx} {\sqrt{(1-\Omega_{0,\Lambda})/\Omega_{0,\Lambda} + x^2} }~.
\end{equation}
The indefinite integral
\begin{equation}\label{eqn:integral}
  \int \frac {dx} {\sqrt{a^2 + x^2}} =
  \ln \, \bigl( x + \sqrt {a^2 + x^2}\bigr) + C~.
\end{equation}
can be found in integral tables or evaluated via the trigonometric
substitution $x = a\,\tan(u)$.  Thus
\begin{equation}\label{eqn:tlapprox}
\boxed{
\frac {t_\mathrm{L}} {t_{\mathrm{H}_0}} \approx
\frac {2}{3 \,\Omega_{0,\Lambda}^{1/2}}
\ln \Biggl[ \frac {1 +  \Omega_{0,\Lambda}^{-1/2}}
{(1 + z)^{-3/2} +
  \sqrt{(1+z)^{-3} + (1 -\Omega_{0,\Lambda})/\Omega_{0,\Lambda}} } \Biggr]} 
\end{equation}
In the limit $z \rightarrow \infty$, $t_\mathrm{L}$ becomes  
the present age of the universe $t_0$:
\begin{equation}\label{eqn:t0approx}
  \boxed{
  \frac {t_0} {t_{\mathrm{H}_0}} \approx  \frac {2}{3 \,\Omega_{0,\Lambda}^{1/2}}
  \ln \Biggl[ \frac {1 + \Omega_{0,\Lambda}^{1/2}}
  {(1 - \Omega_{0,\Lambda})^{1/2}}\Biggr] }~.
\end{equation}
The fractional errors in Equations~\ref{eqn:tlapprox} and
\ref{eqn:t0approx} are $< 10^{-3}$ for all $z$ and all
$\Omega_{0,\mathrm{m}} > 0.1$ only because the high redshifts at which the
omitted $\Omega_{0,\mathrm{r}}(1+z')^4$ term is significant contribute little to
$t_\mathrm{L}$ and $t_0$.  Equation~\ref{eqn:tlapprox} can be used to
calculate accurate  lookback time \emph{differences} $\Delta
t_\mathrm{L} = t_\mathrm{L}(z + \Delta z) - t_\mathrm{L}(z)$ only if
$z \ll 3500$.  The fractional errors in differential lookback times
are $< 10^{-3}$  if $z < 10$ and rise to $10^{-2}$ at $z \sim 60$
and to $10^{-1}$ at $z \sim 500$.

The age of the universe at redshift $z$ was
\begin{equation}
  \frac {t}{t_{\mathrm{H}_0}} \approx \int_z^\infty
   \frac {dz'}
         {(1 + z') [\Omega_{0,\mathrm{m}}(1 + z')^3 + \Omega_{0,\Lambda}]^{1/2}}
\end{equation}
without the radiation term $\Omega_{0,\mathrm{r}}(1+z')^4$.  This
approximation is safe at redshifts $z \lesssim 25$ ($t \gtrsim 0.045
\, t_{\mathrm{H}_0} \sim 6 \times 10^8 \mathrm{~yr}$) because
radiation dominated $E(z)$ for only the first $t \sim 3.4 \times
10^{-6} \, t_\mathrm{H} \sim 5 \times 10^4\,h^{-1} \mathrm{~yr}
\sim 7 \times 10^4 \mathrm{~yr}$ after the big bang
(Section~\ref{sec:times}).
\begin{equation}
  \frac {t} {t_{\mathrm{H}_0}} \approx \int_0^{(1+z)^{-1}} \frac {a^{1/2} \, da}
        { [ (1-\Omega_{0,\Lambda}) + \Omega_{0,\Lambda} a^3]^{1/2}}~.
\end{equation}
Substituting $x \equiv a^{3/2} \Omega_{0,\Lambda}^{1/2}$  yields
the elementary form
\begin{equation}
  \frac {t} {t_{\mathrm{H}_0}} \approx \frac {2} {3\, \Omega_{0,\Lambda}^{1/2}}
  \int_0^{(1+z)^{-3/2}\Omega_\Lambda^{1/2}} \mskip-20mu \frac {dx}
      {[(1-\Omega_{0,\Lambda}) + x^2]^{1/2}}~.
\end{equation}
This is similar to Equation~\ref{eqn:integral}, so
\begin{equation}\label{eqn:tthapprox}
  \boxed{
  \frac {t} {t_{\mathrm{H}_0}} \approx \frac {2} {3\, \Omega_\Lambda^{1/2}} \,
 \ln  \Biggl\{ \biggl( \frac {\Omega_{0,\Lambda}} {1 - \Omega_{0,\Lambda}}
 \biggr)^{1/2} 
 (1+z)^{-3/2} +
 \biggl[ \frac {\Omega_{0,\Lambda}} {(1+z)^3 \,(1-\Omega_{0,\Lambda})} + 1
   \biggr]^{1/2} \Biggr\} }~.
\end{equation}
Equation~\ref{eqn:tthapprox} directly gives the same
$t_0/t_{\mathrm{H}_0}$ as Equation~\ref{eqn:t0approx}, $t/t_{\mathrm{H}_0}$
with fractional errors $< 10^{-2}$ for all $z < 25$, and smaller
errors in time \emph{differences} $\Delta t = -\Delta t_\mathrm{L}$ at
high $z$ than Equation~\ref{eqn:tlapprox}.

\section{Numerical Calculation of Comoving Distance}\label{app:dc}

Equation~\ref{eqn:dc} for comoving distance:
\begin{equation}
D_\mathrm{C} = D_\mathrm{H_0} \int_0^z \frac {dz'}{E(z')}~,
\end{equation}
where
\begin{equation}
   E(z) \equiv [\Omega_{0,\mathrm{m}} (1+z)^3 + \Omega_{0,\Lambda}
     + \Omega_{0,\mathrm{r}} (1+z)^4]^{1/2}~,
\end{equation}
 cannot be expressed in terms of elementary functions.  However, $D_\mathrm{C}$
 varies smoothly with both redshift $z$ (Figure~\ref{fig:dist})
 and normalized matter density $\Omega_{0,\mathrm{m}}$ (Figure~\ref{fig:dcom}), so
 the integral can be approximated by Simpson's rule.  For very large redshifts
 $z \gg 1$, it is more efficient to integrate over $a = (1+z)^{-1}$ instead.
Integrating Equation~\ref{eqn:ddc}:
\begin{equation}
  d D_\mathrm{C} = (c / a) \, dt  
\end{equation}
gives
\begin{equation}
  D_\mathrm{C} = c \int_t^{t_0} \frac {dt}{a} = c \int_a^1
  \frac{1}{a'} \frac {dt}{da'} \,da' =
  c \int_a^1 \frac{1}{{a'}^2} \frac{a'}{\dot{a'}}\,da' =
  c \int_z^1 \frac {da'}{{a'}^2 H} =
  D_{H_0} \int_a^1 \frac{da'}{{a'}^2 E(a')}
\end{equation}
and finally
\begin{equation}\label{eqn:dca}
  D_\mathrm{C} = D_{H_0} \int_a^1 \frac{da'}
  {(\Omega_{0,\mathrm{m}}\, a' + \Omega_{0,\Lambda} \, {a'}^4 + \Omega_{0,\mathrm{r}})^{1/2}}~.
\end{equation}
The FORTRAN function {\tt dcmpc} below evaluates
Equation~\ref{eqn:dca} to return an accurate $D_\mathrm{C}$ (in Mpc)
for redshifts $z$ even into the photon-dominated era $z > z_\mathrm{eq} \approx 3500$, given
$h \sim 0.7$ and $\Omega_{0,\mathrm{m}} \sim 0.3$.  It can be copied and pasted directly into
a text editor such as Emacs.  The corresponding Python function {\tt dc.py} is available
at {\tt https://github.com/allison-matthews/astro-cosmo}

\vfil



{\tt
\noindent C ********************************************************************** 
\newline \hphantom{~~~~~~}function dcmpc (z, h, Omega0m)
\newline C **********************************************************************
\newline \hphantom{~~~~~~}real*8 aofz, da, simp
\newline \hphantom{~~~~~~}aofz = z
\newline \hphantom{~~~~~~}aofz = 1.~/ (1.~+ aofz)
\newline \hphantom{~~~~~~}Omega0r = 4.2e-05 / h**2
\newline \hphantom{~~~~~~}Omega0l = 1.~- Omega0m - Omega0r
\newline \hphantom{~~~~~~}dh0mpc = 2997.92458 / h
\newline C Evaluate Condon \& Matthews 2018, PASP, Eq.~B5 using Simpson's rule:
\newline C simp = (da / 3.)~* (y0 + 4.*y1 + 2.*y2 + 4.*y3 + 2.*y4 + ...~+ yn)
\newline C where n must be even
\newline \hphantom{~~~~~~}n = 10
\newline \hphantom{~~~~~~}if (z > 0.1) n = 100
\newline \hphantom{~~~~~~}if (z > 3.0) n = 1000
\newline \hphantom{~~~~~~}if (z > 90.)~n = 10000
\newline \hphantom{~~~~~~}if (z > 2700.)~n = 100000
\newline \hphantom{~~~~~~}da = n
\newline \hphantom{~~~~~~}da = (1.~- aofz) / da
\newline \hphantom{~~~~~~}simp = 0.0
\newline \hphantom{~~~~~~}do 100 isub = 0, n
\newline \hphantom{~~~~~~~~~}a = isub
\newline \hphantom{~~~~~~~~~}a = a * da + aofz
\newline \hphantom{~~~~~~~~~}y = 1.~/ sqrt (Omega0m * a + Omega0l * a**4 + Omega0r)
\newline C     test for even or odd y subscript and assign weighting factor
\newline \hphantom{~~~~~~~~~}itest = (isub / 2) * 2
\newline \hphantom{~~~~~~~~~}if (itest == isub) factor = 2.
\newline \hphantom{~~~~~~~~~}if (itest /= isub) factor = 4.
\newline \hphantom{~~~~~~~~~}if (isub == 0) factor = 1.
\newline \hphantom{~~~~~~~~~}if (isub == n) factor = 1.
\newline \hphantom{~~~~~~~~~}simp = simp + factor * y
\newline \hphantom{~}100\hphantom{~~~~~}continue
\newline \hphantom{~~~~~~}dcmpc = dh0mpc * (da / 3.)~* simp
\newline \hphantom{~~~~~~}return
\newline \hphantom{~~~~~~}end
}


\vfil\eject

\section{Luminosity Functions, Source Counts, and Sky Brightness}\label{app:s2ns}

Let $N(>S)$ be the number of sources per steradian stronger than flux
density $S \equiv F_\nu$, $n (S) \equiv -dN / dS$ be the differential
source count, and $\eta (S,z) \,dS \,dz$ be the number of sources per
steradian with flux densities $S$ to $S+dS$ in the redshift range $z$
to $z + dz$.  The spectral luminosity function $\rho (L_\nu \,\vert\,
z)\, dL$ is the comoving number density of sources at a given redshift
$z$ having spectral luminosities $L_\nu$ to $L_\nu + dL_\nu$, and $d
V_\mathrm{C}$ is the comoving volume element covering $\omega = 1
\mathrm{~sr}$ of sky between $z$ and $z +dz$.  The number of sources
equals the comoving density times the comoving volume:
\begin{equation}
  \eta (S,z) \, dS \, dz = \rho(L_\nu \,\vert\, z) \,dL \, d V_\mathrm{C}~.
\end{equation}
For sources with spectral indices $\alpha$,
\begin{equation}
  L_\nu = 4 \pi D_\mathrm{L}^2 \,(1+z)^{-1-\alpha}\, F_\nu =
  4 \pi D_\mathrm{C}^2\, (1+z)^{1-\alpha} \,S
\end{equation}
(Equation~\ref{eqn:fluxdensity})
and 
\begin{equation}
  dV_\mathrm{C} = \frac {D_\mathrm{C}^2 D_\mathrm{H_0}} {E(z)} \,dz
\end{equation}
(Equation~\ref{eqn:dvc}).  Thus
\begin{equation}
  \eta (S,z)\,dS\,dz  = \rho(L_\nu \,\vert\, z) \,
  4 \pi D_\mathrm{C}^2\, (1+z)^{1-\alpha} \,dS
\, \frac {D_\mathrm{C}^2 D_\mathrm{H_0}} {E(z)}dz~.
\end{equation}
Multiplying both sides by
\begin{equation}
  S^2 = \biggl[\frac {(1+z)^{\alpha-1}\,L_\nu }{4 \pi D_\mathrm{C}^2}\biggr]^2
\end{equation}
gives
\begin{equation}
  S^2 \eta (S,z) = L_\nu^2 \, \rho (L_\nu \,\vert\, z) \,
  \biggl[ \frac {(1+z)^{\alpha-1} D_\mathrm{H_0}} {4 \pi E(z)} \biggr]~.
\end{equation}
Frequently the spectral luminosity function is specified as the
density of sources per decade of luminosity
\begin{equation}
  \rho_\mathrm{dex} (L_\nu \,\vert\, z) =
  \ln(10)  L_\nu \, \rho (L_\nu \,\vert\, z)~.
\end{equation}
The luminosity-weighted spectral luminosity function 
\begin{equation}\label{eqn:udex}
  U_\mathrm{dex}(L_\nu \,\vert\, z) \equiv L_\nu \,
  \rho_\mathrm{dex} (L_\nu \,\vert\, z)
  = \ln(10)  L_\nu^2  \rho(L_\nu \,\vert\, z)
\end{equation}
(SI units $\mathrm{W~Hz}^{-1} \mathrm{~m}^{-3} \mathrm{~dex}^{-1} =
\mathrm{J~m}^{-3} \mathrm{~dex}^{-1}$, the same as energy density)
emphasizes the luminosity ranges contributing the most to the spectral
luminosity density.  In terms of these quantities,
\begin{equation}
  S^2 \eta (S, z) = U_\mathrm{dex} (L_\nu \,\vert\, z) \,
  \biggl[ \frac {(1+z)^{\alpha-1} D_\mathrm{H_0}}{4 \pi \ln(10) E(z)} \biggr]
\end{equation}
and the total brightness-weighted source count is obtained by
integrating over all redshifts:
\begin{equation}\label{eqn:s2ns}
\boxed{
  S^2 n(S) = \frac {D_\mathrm{H_0}} {4 \pi \ln(10)} \int_0^\infty
  U_\mathrm{dex} (L_\nu \,\vert\, z)\,
  \biggl[ \frac {(1 + z)^{\alpha-1}} {E(z)} \biggr] \, dz }~.
\end{equation}
Note that $S^2 n(S)$ has dimensions of spectral brightness (SI units
$\mathrm{W~m}^2 \mathrm{~Hz}^{-1} \mathrm{~sr}^{-1}$ or astronomically
practical units $\mathrm{Jy~sr}^{-1}$).

\vskip 10pt


\end{document}